\documentclass[pr, aps, 10pt, twocolumn, preprintnumbers, showpacs, superscriptaddress]{revtex4-1}

\usepackage[T1]{fontenc}
\usepackage{amssymb}
\usepackage{graphicx}
\usepackage{amsmath}
\usepackage{slashed}
\usepackage{tensor}
\usepackage{epstopdf}
\usepackage{extarrows}
\usepackage{xcolor}
\usepackage{tikz}
\usepackage{mathrsfs,amsthm,latexsym,amsfonts,txfonts,cancel,color,enumerate}
\usepackage{bm}
\usepackage{fancyhdr}
\usepackage{hyperref}

\begin{document}
\title{The shadows and photon rings of two minimal deformations of Schwarzschild black holes}

\author{Hong-Er Gong}
\email{gonghonger@yeah.net}
\affiliation{Guangxi Key Laboratory for Relativistic Astrophysics, School of Physical Science and Technology, Guangxi University, Nanning 530004, China}
\author{Junlin Qin}
\affiliation{Guangxi Key Laboratory for Relativistic Astrophysics, School of Physical Science and Technology, Guangxi University, Nanning 530004, China}
\author{Yusen Wang}
\affiliation{Guangxi Key Laboratory for Relativistic Astrophysics, School of Physical Science and Technology, Guangxi University, Nanning 530004, China}
\author{Bofeng Wu}
\email{wubofeng@gxu.edu.cn}
\affiliation{Guangxi Key Laboratory for Relativistic Astrophysics, School of Physical Science and Technology, Guangxi University, Nanning 530004, China}
\author{Zhan-Feng Mai}
\email{zf1102@gxu.edu.cn}
\affiliation{Guangxi Key Laboratory for Relativistic Astrophysics, School of Physical Science and Technology, Guangxi University, Nanning 530004, China}
\author{Sen Guo}
\affiliation{College of Physics and Electronic Engineering, Chongqing Normal University, Chongqing 401331, China}
\author{Enwei Liang}
\email{lew@gxu.edu.cn}
\affiliation{Guangxi Key Laboratory for Relativistic Astrophysics, School of Physical Science and Technology, Guangxi University, Nanning 530004, China}

\begin{abstract}
This paper primarily investigates the optical characteristics of two minimal Schwarzschild black hole deformations, the Kazakov-Solodukhin and Ghosh-Kumar black holes, under different accretion models.
The event horizon, photon sphere, and critical impact parameter of the former increase compared with the Schwarzschild black hole, but those of the latter decrease.
The data from the Event Horizon Telescope Collaboration are used to constrain the parameter ranges of the two black holes.
In the case of spherical accretion, the quantum-correction of Kazakov-Solodukhin black hole leads to the increase of black hole shadow size and the decrease of integrated intensity, while the shadow size of magnetically charged Ghosh-Kumar black hole decreases and the integrated intensity increases.
The shadow radius of the black hole is independent of the spherical accretion models.
For an optically and geometrically thin accretion disk, the integrated intensity is mainly contributed by direct emission, and the contributions of photon rings and lensed rings are very small.
In addition, the photon rings and lensed rings of Kazakov-Solodukhin black hole are narrower, while those of Ghosh-Kumar black hole are wider.
Whereas the Kazakov-Solodukhin black hole exhibits higher brightness, the Ghosh-Kumar black hole shows lower brightness.
Additionally, a disk closer to the black hole correlates with a smaller shadow radius.
This paper proposes a method to distinguish different black holes in a specific thin disk model.
\end{abstract}

\maketitle
\flushbottom

\noindent

\section{Introduction\label{sub1}}
In 1915, Albert Einstein proposed General Relativity (GR).
In 1916, Karl Schwarzschild derived the first exact solution to Einstein's equations, known as the Schwarzschild (SC) solution \cite{r1}.
Later, it was recognized as a black hole (BH) solution \cite{r2,r3}, which has since attracted extensive attention.
In 2015, the Laser Interferometer Gravitational Wave Observatory (LIGO) successfully detected gravitational wave signals radiated by the merger of binary BHs.
The discovery of stellar-BH binary systems through radial velocity measurements indirectly confirmed the existence of BHs in the Universe  \cite{r4,r5,r6,r7}.
More direct evidence for the existence of BHs come from the ultra-high angular resolution images of M87* and Sgr A* released by the Event Horizon Telescope Collaboration (EHT) \cite{r8,r9,r10,r11,r12,r13,r14,r15,r16,r17,r18,r19,r20,r21,r22,r23,r24}.
Due to the gravitational lensing effect \cite{r25,r26,r27,r28}, light is deflected near BHs, leading to a decrease in the observed intensity within the boundary.
Therefore, a region with lower brightness, known as the BH shadow, can be observed on the distant observation plane.
The shadow boundary is defined by escaping photons, not the event horizon itself. Consequently, the observed shadow exceeds the event horizon in size.
Furthermore, the BH shadow is closely related to spacetime geometry, and thus serves as a powerful tool for estimating BH parameters \cite{r29,r30,r31,r32} and testing GR or other modified gravity theories~\cite{r33,r34,r35,r36,r37,r38,r39}.

In recent years, this research field has been further expanded to a series of important frontier topics, including the multi-photon sphere structures of BHs~\cite{d1}, the influence of magnetic field effects on the morphology and observational features of BH shadows~\cite{d2}, and the constraints on compact extra dimension theories from BH shadow observations~\cite{d3}. These studies further enrich the physical connotation of BH shadow observations, and provide more abundant theoretical tools for testing gravitational theories in the strong-field regime.

In astrophysics, most BHs are rotating because they form at the final stage of the collapse of massive stars.
Bardeen was the first to analytically obtain the shadow of the Kerr BH and demonstrated that it is D-shaped~\cite{r40}.
The shadows of rotating BHs, as well as the relationships between shadow shapes, spin parameters, and observation angles, were further studied~\cite{r41,r42}.
In addition, the shadows of rotating BHs in modified gravity theories have also attracted public attention \cite{r43,r44}.
In the theory of Einstein-Maxwell-Dilation (EMD) gravity, Chen et al. explored the shadows of rotating EMD BHs at different inclination angles \cite{r45}.
It should be emphasized that the rotational energy of a rotating BH can be extracted from the ergosphere through the Penrose process or the Blandford and Znajek mechanism, leading to the degradation of the rotating BH into a non-rotating one \cite{r46,r47}.
It can also be demonstrated that the influence of spin on the BH shadow radius is minimal \cite{r39}, especially at low spin values.
Therefore, the shadows and optical images of static spherically symmetric BHs likewise require detailed investigation to explore their fundamental properties.

In studies on static spherical BH imaging, Bambi investigated the apparent images of SC BHs and static wormholes illuminated by optically thin spherical accretion flows.
He pointed out that based on the shadow images, the former can be relatively easily distinguished from the latter \cite{r48}. 
Other researchers have also discussed the differences in shadows between non-rotating BHs and wormholes under spherical accretion conditions~\cite{r49}.
Narayan et al. studied the observed shadows of spherically accreting SC BHs and argued that the shadow radius is independent of the details of spherical accretion \cite{r50}.
Wu et al. obtained more comprehensive observed shadows of SC BHs using analytical methods \cite{r51}.
Zeng et al. investigated the shadows and photon rings of four-dimensional (4D) Gauss-Bonnet BHs, proposing that changes in the spherical accretion model or the emissivity profile do not alter the shadow size but significantly affect its luminosity~\cite{r52}.

In the study of BH images, spherical accretion provides an idealized model for the accretion mechanism near astrophysical BHs.
Due to the strong gravity near BHs, matter in the universe is captured and forms a huge disk-like accretion flow around them.
For distant observers, radiation from the accretion disk illuminates the spacetime around the BH, enabling the observation of its shadow.
Building on Luminet's foundational study of rotating, geometrically thick accretion disks \cite{r53.5}, Gralla et al. conducted the first systematic analysis of the shadow cast by a SC BH surrounded by an optically and geometrically thin accretion disk.
They specifically demonstrated that such shadows exhibit not only a photon ring but also a lensed ring structure.
However, due to the insufficient angular resolution of the EHT, these two rings have almost no observational effect under the specific emission profile of the accretion disk \cite{r53}.
Since then, more and more researchers have started to focus on the influence of thin accretion disks on the observed BH shadows \cite{r54,r55,r56,r57}.

A minimal deformation of the SC BH refers to modifying the metric of the classical SC BH in the minimal way, while preserving its core geometric and physical properties, such as spherical symmetry, asymptotic flatness, and features of the event horizon, but also introducing new physical features. 
When the deformation parameter is set to zero, the solution reverts to the SC metric~\cite{r39}. 
Such deformations can, to some extent, be used to construct regular BHs that overcome issues like the singularity problem. 
This paper studies two minimal deformations of the SC BH: the Kazakov–Solodukhin (KS) BH originating from quantum-corrections~\cite{r58}, and the Ghosh–Kumar (GK) BH arising from nonlinear electrodynamics (NED)~\cite{r59}. 
In both cases, a minimal deformation of the SC BH is achieved through a simple algebraic replacement of the radial coordinate in the metric function. 
In this way, a deformation ``hair'' parameter is introduced into the SC BH for each case. 
Interestingly, such deformation ensures the removal of the central singularity of the SC BH. 
To distant observers, the KS BH and GK BH closely resemble the SC BH, sharing the same ADM mass and having nearly identical event horizon positions. 
It is worth noting that the expressions for the event horizons of the KS BH and the GK BH are remarkably similar, differing only in sign (see Eqs.~\eqref{rhks} and~\eqref{rhgk}). 
This implies that the two hair parameters influence the deformation of the BH through formally similar mechanisms.

Compared to the SC BH, quantum-corrected BHs generally possess larger event horizons, photon spheres, and photon ring radii, along with lower effective potentials for photons. 
This is a widely held conjecture in the community~\cite{r44,b1,b2}. 
In contrast, typical BHs in NED theories exhibit the opposite characteristics relative to the SC BH: their event horizons, photon spheres, and photon ring radii are all smaller, while the effective potentials for photons are higher~\cite{b3,r39,b4}. 
Existing studies predominantly address the optical characteristics of quantum-corrected BHs or NED BH in isolation. 
Few works have conducted a precise, one-to-one comparative study under identical accretion models and observational frameworks for the two classes of SC type minimally deformed BHs that exhibit symmetric forms but possess completely opposite physical origins.
The similarity between the KS BH and the GK BH offers a natural opportunity to compare these two opposing theoretical properties.
The core objective of this paper is to elucidate the opposing regulatory mechanisms of quantum-corrections and NED effects on BH shadows and photon rings, using the representative pair of minimally deformed models, KS and GK. 
Based on this objective, this study selects this highly representative pair of models to perform quantitative investigations within a unified ray-tracing, radiative transfer, and imaging framework, aiming to provide observable criteria for distinguishing between these two classes of BH deformations originating from different physics.
We conjecture that this contrast manifests as opposite behaviors in the characteristic parameters of the shadow images for the two BHs. 
In fact, the event horizon behavior described above serves as an initial example supporting our conjecture. 
This difference can also be reflected in the asymptotic behaviors of the metric functions at large $r$ when the second order terms are considered (see the opposite signs of the third terms in Eqs.~\eqref{jiks} and~\eqref{jigk}).

Bolokhov et al. first investigated the outburst of overtones and Hawking evaporation of KS BHs, showing that for sufficiently small BHs, the changes in Hawking radiation and its Quasi-Normal modes (QNMs) overtone frequencies are more pronounced in the final stage of BH evaporation \cite{r60}.
Separately, Konoplya studied the basic geometric features of KS BHs, including QNMs, scattering, and shadows, and found that the shadow radius decreases~\cite{r61}. 
Peng et al. discovered that KS BHs violate the universal inequalities proven for asymptotically-flat BHs satisfying the null energy condition in Einstein gravity. 
They also studied the shadows, lensed rings, and photon rings of KS BHs~\cite{r44}.
Guo et al. investigated the impact of quantum-corrections on polarization information in KS BHs, demonstrating that larger quantum deformations lead to the expansion of the polarized region and a reduction in polarization intensity \cite{r63}.
Furthermore, Huang et al. showed that the observable properties of KS BHs depend not only on the accretion disk shape but also on spacetime characteristics and the observer's inclination angle. 
Moreover, these properties differ between BHs surrounded by optically thick and thin accretion disks~\cite{r64}.
Moreover, Karmakar compared the behavior of GK BHs and Kerr-Newman (KN) BHs under scalar field scattering~\cite{r65}.
Yang et al. revealed that the Einstein ring of GK BHs exhibits a transition from an axisymmetric closed circle to an arc shape, inducing deformations in the shadow's morphology~\cite{r66}.

However, these studies have focused on inclined disks and rotational profiles of KS and GK BH shadows, whereas the influence of spherical accretion and face-on thin disks remains unclear.
Since accretion flows play a crucial role in BH shadow observations, we primarily investigate the radiation intensity and optical images of BHs under two distinct accretion models: spherical accretion and face-on disk accretion.
By analyzing the shadows and observational characteristics of three different BHs and leveraging the interaction of different BH parameters, we provide a method to distinguish between KS, GK, and SC BHs.
Furthermore, through BH images, the present study also validates the aforementioned conjecture regarding quantum-corrections and NED BHs, revealing the influence mechanisms of different types of minimal deformations on BH shadows. 
Moreover, our findings confirm the similarity in how KS BHs and GK BHs affect BH shadows, offering new perspectives on understanding the role of BH hair in shaping shadow characteristics. 
Therefore, investigating such BHs is of great significance for understanding the universe.

This paper is organized as follows. In Sec.~\ref{sub2}, we derive the metrics for the two minimal deformations of the SC BH from their respective theoretical actions.
In Sec.~\ref{sub3}, we study the light deflection and critical impact parameters for KS and GK BHs by analyzing the null geodesics and effective potentials. 
We also present the constraints on the BH parameters derived from EHT observational data.
In Sec.~\ref{sub4}, we explore the specific intensities and optical appearances of KS and GK BHs under two spherical accretion models: static and infalling.
In Sec.~\ref{sub5}, we then investigate the observational signatures of these two BHs surrounded by optically and geometrically thin accretion disks.
Finally, we summarize and discuss our results in Sec.~\ref{sub6}.
In this paper, we adopt geometric units, i.e., $c=G=1$.

\section{Introduction to KS and GK BHs\label{sub2}}
Consider the following form of the static spherically symmetric metric, to which all subsequent metric solutions conform:
    \begin{equation}
        \mathrm{d}s^2=-f(r)\mathrm{d}t^2+\dfrac{1}{f(r)}\mathrm{d}r^2+r^2(\mathrm{d}\theta^2+\sin^2{\theta}\mathrm{d}\varphi^2)\label{1}
    \end{equation}
with $f(r)$ being the metric function.

\subsection{KS metric\label{sub2.1}}

The KS metric is a spherically symmetric deformation of the SC BH due to quantum fluctuation corrections. Kazakov and Solodukhin showed in reference~\cite{r58} that when considering the spherically symmetric quantum fluctuations of the SC BH, the Einstein–Hilbert action 
\begin{equation*}
    \mathscr{S}=\dfrac{1}{16\pi G}\int\mathrm{d}x^4\sqrt{-g}R,
\end{equation*}
can be reduced to a 2D effective dilaton action by integrating out the two angular coordinates,
    \begin{equation}
        \mathscr{S}_1 =\dfrac{1}{8}\int\mathrm{d}^2z\sqrt{-g}\left[\xi^2R^{(2)}-2(\nabla\xi)^2+\dfrac{2}{G}U(\xi)\right]\, , \label{19}
    \end{equation}
where $\xi=r/\sqrt{G}$ is the dimensionless radial coordinate, $R^{(2)}$ is the 2D Ricci scalar,$G$ is Newton's constant, and $U(\xi)$ is the 2D dilaton potential. The detailed derivation is provided in appendix~\ref{app}, from Eq.~\eqref{a.1} to Eq.~\eqref{a.8}.
Variation of the action~\eqref{19} with respect to the 2D metric $g^{(2)\alpha\beta}$ and $\xi$ yields the field equations
    \begin{equation}
        \begin{split}
        &2\xi\nabla_{\alpha}\nabla_{\beta}\xi=g^{(2)}_{\alpha\beta}\left[\dfrac{1}{G}U(\xi)+2\xi\square\xi+(\nabla\xi)^2\right],\\[1em]
            &0=2\square\xi+\xi R^{(2)}+\dfrac{1}{G}\dfrac{\mathrm{d}U(\xi)}{\mathrm{d}\xi}.
        \end{split}\label{20}
    \end{equation}
According to the derivation from Eq.~\eqref{AE} to Eq.~\eqref{AKS} in appendix~\ref{app}, the solution to this field equation is
    \begin{equation}
        f(r)=-\dfrac{2M}{r}+\dfrac{1}{r}\int^rU(r)\mathrm{d}r,\label{x4}
    \end{equation}
where the integral is an indefinite integral without the integration constant, and $M$ stands for the BH mass. It is straightforward to see that when $U(r)=1$, the metric~\eqref{1} reduces to the SC metric.

Under the assumption of no non-spherical metric deformations, Kazakov and Solodukhin proposed a potential function $U(r)$ \cite{r58}, which allows both an analytical solution of the field equations and ensures the renormalizability of the 2D dilaton theory. This function satisfies the renormalization group equation
\begin{equation}
    \dfrac{\partial U(\psi,c)}{\partial c}=\dfrac{1}{\psi}U(\psi,c)-2\dfrac{\partial U(\psi,c)}{\partial\psi},\label{x1}
\end{equation}
where $c=\ln(\mu/\mu_0)$, $\mu$ being a scale parameter, $\psi=\xi^2/8$. The general solution of the renormalization group equation satisfying the ``initial'' condition $U(\psi,c=0)=1$ is
\begin{equation}
    U(\psi,c)=\sqrt{\psi}g(\psi-2c),
\end{equation}
where $g(\psi-2c)$ is an arbitrary function chosen from the initial condition. It is noteworthy that the region we discuss is $\{\psi\geq0,c\geq0\}$. Under the ``initial'' condition $U(\psi,c=0)=1$ and boundary condition $U(\psi=0,c)=0$, $g(\psi)=1/\sqrt{\psi}$, thus it is easy to verify that the solution of Eq.~\eqref{x1} has the form
\begin{equation}
    U(\psi,c)=\begin{cases}
        0,&\text{if}\quad\psi\leq2c,\\[1em]
        \dfrac{\sqrt{\psi}}{\sqrt{\psi-2c}},&\text{if}\quad\psi>2c.
    \end{cases}\label{x2}
\end{equation}
Substituting $\psi=r^2/8G$ into Eq.~\eqref{x2} yields
\begin{equation}
    U(r,c)=\begin{cases}
        0,&\text{if}\quad0\leq r\leq 4\sqrt{cG},\\[1em]
        \dfrac{r}{\sqrt{r^2-16cG}},&\text{if}\quad r>4\sqrt{cG}.
    \end{cases}\label{x3}
\end{equation}
Setting $a^{\mathrm{KS}}=4\sqrt{cG}$, and substituting Eq.~\eqref{x3} into Eq.~\eqref{x4} give the expression for the KS metric function (we have used the geometric units with $G=1$)
\begin{equation}
        f^{\mathrm{KS}}(r)=\dfrac{\sqrt{r^2-(a^{\mathrm{KS}})^2}}{r}-\dfrac{2M}{r},\label{ks}
\end{equation}
where $r\geq a^\mathrm{KS}$. Compared to the SC BH, quantum deformation shifts the central singularity to a finite radius and characterizes a universal ``hair'' structure \cite{r39}. From a physical perspective, the singularity is smeared over a 2D sphere. The asymptotic behavior of the above expression at large $r$ is given by
\begin{equation}
    f^\mathrm{KS}(r)\approx1-\dfrac{2M}{r}-\dfrac{(a^{\mathrm{KS}})^2}{r^2}.\label{jiks}
\end{equation}

\subsection{GK metric}
We next introduce another minimal deformation of the SC BH caused by the minimal coupling of GR with NED~\cite{r59}. The action of the NED theory is given as
    \begin{equation}
        \mathscr{S}_2=\int \mathrm{d}^{4}x\sqrt{-g}\left(\frac{1}{16\pi}R-\frac{1}{4\pi}\mathscr{L}(F)\right),\label{22}
    \end{equation}
where $R$ is the Ricci scalar, $g$ is the determinant of the metric tensor. $\mathscr{L}(F)$ is the Lagrangian density of NED with
    \begin{equation}
        F=\dfrac{1}{4}F^{\mu\nu}F_{\mu\nu},
    \end{equation}
where $F_{\mu\nu}=\partial_{\mu}A_{\nu}-\partial_{\nu}A_{\mu}$ is the field strength tensor. By varying the action \eqref{22}, the field equations are derived as
    \begin{equation}
    \begin{aligned}
        &R_{\mu\nu}-\dfrac{1}{2}Rg_{\mu\nu}=2\left(\frac{\partial \mathscr{L}(F)}{\partial F} F_{\mu \lambda} F_{\nu}^{\ \lambda}-g_{\mu \nu} \mathscr{L}(F)\right),
        \\[1em]
        &\nabla_\mu\left(\dfrac{\partial\mathscr{L}(F)}{\partial F}F^{\mu\nu}\right)=0,\qquad\nabla_\mu(^*F^{\mu\nu})=0,\label{23}
    \end{aligned}
    \end{equation}
where $^{*}$ denotes the Hodge dual.
Considering the field strength tensor defined as
\begin{equation}
F_{\mu\nu}=2\delta_{[\mu}^{\theta}\delta_{\nu]}^{\varphi}a(r)\sin\theta,
\end{equation}
substitution into $\nabla_\mu(^*F^{\mu\nu})=0$ in Eq.~\eqref{23} then yields
\begin{equation}
    \dfrac{\mathrm{d}a(r)}{\mathrm{d}r}\sin\theta\mathrm{d}r\wedge\mathrm{d}\theta\wedge\mathrm{d}\varphi = 0\, ,
\end{equation}
so $a(r)\equiv a^{\mathrm{GK}}=\text{const}$, $F_{\theta\varphi}=a^{\mathrm{GK}} \sin{\theta}$, and $F=(a^{\mathrm{GK}})^2/2r^4$, where $a^{\mathrm{GK}}$ corresponds to the magnetic charge.
We adopt the Lagrangian density proposed by Ghosh and Walia \cite{r59}
    \begin{equation}
        \mathscr{L}(F)=\frac{4MF^{5/4}}{\sqrt{a^{\mathrm{GK}}}(\sqrt{2}+2a^{\mathrm{GK}}\sqrt{F})^{3/2}}.
    \end{equation}
Analogously, $M$ stands for the BH mass. The field equations~\eqref{23} provides the GK BH metric function
    \begin{equation}
        f^{\mathrm{GK}}(r)=1-\dfrac{2M}{\sqrt{r^2+(a^{\mathrm{GK}})^2}}.\label{gk}
    \end{equation}
It should be emphasized that the regularity of this BH needs to be clarified in detail. The metric function of the GK BH is completely regular, smooth and non-singular throughout the spacetime with $r>0$, including at the event horizon, which solves the coordinate singularity problem of the classical SC BH at the event horizon. For this reason, this solution is widely referred to as a regular BH solution with minimal deformation of the SC BH in existing literature~\cite{r39}. However, it should be clearly stated that the GK BH is not globally regular in the strict curvature sense: there is still a scalar polynomial singularity at the origin $r=0$, which can be verified by calculating the Kretschmann scalar and Ricci scalar of the spacetime. In addition, this spacetime exhibits an interesting reflection symmetry $r\to-r$~\cite{r39,r59}.
The asymptotic behavior of the above expression at large $r$ is given by
\begin{equation}
    f^\mathrm{GK}(r)\approx1-\dfrac{2M}{r}+\dfrac{(a^\mathrm{GK})^2M}{r^3}.\label{jigk}
\end{equation}

It is worth noting that for vanishing deformation radius and magnetic charge, namely $a^{\mathrm{KS}}\to0$ and $a^{\mathrm{GK}}\to0$, both of the above metrics reduce to the SC metric. This preserves the fundamental properties of the SC solution, including spherical symmetry (in the absence of rotation) and asymptotic flatness.
Therefore, they can be viewed as minimal deformations of the SC BH \cite{r39}.

To explore the optical properties of these BHs and to investigate how different deformation induced hair parameters and their interactions with spacetime affect the resulting images, we treat the key parameters in each metric, namely the deformation parameter $a^{\mathrm{KS}}$  for the KS BH and the magnetic charge $a^{\mathrm{GK}}$ for the GK BH as distinct forms of BH hair.

It should be particularly noted that although both $a^\mathrm{KS}$ of KS BH and $a^\mathrm{GK}$ of GK BH are parameters of minimal deformation of the SC spacetime, they have completely independent physical origins. The former arises from the quantum renormalization correction of the 2D dilaton theory after dimensional reduction of the 4D Einstein-Hilbert action, serving as a deformation parameter characterizing quantum fluctuations of spacetime. The latter originates from the minimal coupling of GR and NED, representing a physical quantity that characterizes the intrinsic magnetic charge of the BH. They only exhibit symmetric minimal deformation features in the algebraic form of the metric, with no equivalence in their physical connotations.

\section{Light deflection, effective potential, and impact parameter\label{sub3}}
For SC BH, KS BH, and GK BH, their metric functions $f(r)$ in the metric \eqref{1} have been given in Sec.~\ref{sub2}.
The event horizons of the three BHs read
\begin{itemize}
    \item SC BH
\end{itemize}
    \begin{equation}
        r^{\mathrm{SC}}_{h}=2M,
    \end{equation}
 \begin{itemize}
     \item KS BH
 \end{itemize}
    \begin{equation}
        r^{\mathrm{KS}}_{h}=\sqrt{4M^2+{(a^{\mathrm{KS}})^2}},\label{rhks}
    \end{equation}
\begin{itemize}
    \item GK BH
\end{itemize}
    \begin{equation}
        r^{\mathrm{GK}}_{h}=\sqrt{4M^2-{(a^{\mathrm{GK}})^2}}.\label{rhgk}
    \end{equation}
For the KS BH, there is always an event horizon regardless of the parameter {$a^\mathrm{KS}$}, with the horizon radius satisfying $r^{\mathrm{KS}}_{h}>r^{\mathrm{SC}}_{h}$.
For the GK BH, when $(a^\mathrm{GK})^2 \geq 4M^2$, the event horizon disappears. Specifically, when $(a^\mathrm{GK})^2 = 4M^2$, the GK BH becomes an extreme BH, and when $(a^\mathrm{GK})^2 > 4M^2$, it becomes a naked singularity.
An event horizon satisfying $r^{\mathrm{GK}}_{h}<r^{\mathrm{SC}}_{h}$ is only attainable when ${(a^{\mathrm{GK}})^2}<4M$.
It is important to emphasize that in static spherically symmetric spacetimes, whether for extreme BHs, naked singularities, or even wormholes and boson stars, the presence of a photon sphere determines the shadow's image \cite{r37}.
Thus, as long as the mass of a celestial body is sufficient to generate a photon sphere, the existence of a shadow can be observed.
In the following, we primarily discuss the non-extreme GK BHs case.

Our analysis centers on light ray trajectories, necessitating examination of null geodesics originating from BH accretion flows.
Without loss of generality, the following analysis considers photon geodesics in the equatorial plane, i.e., taking $\theta=\pi/2$. For the spherically symmetric metric in Eq.~\eqref{1}, there are two Killing vector fields $\xi_1^\mu$ and $\xi_2^\mu$, given by
    \begin{equation}
        \xi_1^\mu=\delta^\mu_t,\qquad \xi_2^\mu=\delta^\mu_\varphi.
    \end{equation}
Let the photon's 4-momentum be $K^\mu=\mathrm{d}x^\mu/\mathrm{d}\lambda$. Along the null geodesic, there are two conserved quantities: the photon's energy $E$ and angular momentum $L$, i.e.,
    \begin{equation}
        E=-g_{\mu\nu}\xi_1^\mu\dfrac{\mathrm{d}x^\nu}{\mathrm{d}\lambda}=f(r)\dfrac{\mathrm{d}t}{\mathrm{d}\lambda},\quad L=-g_{\mu\nu}\xi_2^\mu\dfrac{\mathrm{d}x^\nu}{\mathrm{d}\lambda}=r^2\dfrac{\mathrm{d}\varphi}{\mathrm{d}\lambda}.
    \end{equation}
The 4-momentum of the photon also satisfies $g_{\mu\nu}K^\mu K^\nu=0$, so the components of the photon's 4-momentum are
    \begin{equation}
        K^\mu=\left(\dfrac{E}{f(r)},\pm\sqrt{E^2-\dfrac{f(r)L^2}{r^2}},0,\dfrac{L}{r^2}\right),\label{17}
    \end{equation}
where the positive sign corresponds to moving away from the BH, and the negative sign corresponds to approaching the BH.
Meanwhile, the differential equation for the radial coordinate with respect to the affine parameter is derived as
    \begin{equation}
        \left(\dfrac{\mathrm{d}r}{\mathrm{d}\lambda}\right)^2=E^2-f(r)\dfrac{L^2}{r^2}.\label{5}
    \end{equation}

The impact parameter and the effective potential of the photon are defined by
    \begin{equation}
        b:=\dfrac{L}{E},\qquad V_{\mathrm{eff}}:=\dfrac{1}{b}.
    \end{equation}
Next, we discuss the effective potential for circular orbits. For a circular orbit, $\mathrm{d}r/\mathrm{d}\lambda=0$. Substituting this into Eq.~\eqref{5}, we have
    \begin{equation}
        V_{\mathrm{eff}}=\dfrac{\sqrt{f(r)}}{r}.
    \end{equation}
For the three metrics under consideration, the effective potentials are respectively
\begin{itemize}
    \item SC BH
\end{itemize}
    \begin{equation}
        V^{\mathrm{SC}}_{\mathrm{eff}}=\sqrt{\dfrac{r-2M}{r^3}},
    \end{equation}
\begin{itemize}
    \item KS BH
\end{itemize}
    \begin{equation}
        V^{\mathrm{KS}}_{\mathrm{eff}}=\sqrt{\dfrac{\sqrt{r^2-{(a^{\mathrm{KS}})^2}}-2M}{r^3}},
    \end{equation}
\begin{itemize}
    \item GK BH
\end{itemize}
    \begin{equation}
        V^{\mathrm{GK}}_{\mathrm{eff}}=\left(\dfrac{\sqrt{r^2+{(a^{\mathrm{GK}})^2}}-2M}{r^2\sqrt{r^2+{(a^{\mathrm{GK}})^2} }}\right)^{1/2}.
    \end{equation}

\begin{figure}[t]
        \centering
        \includegraphics[width=1\linewidth]{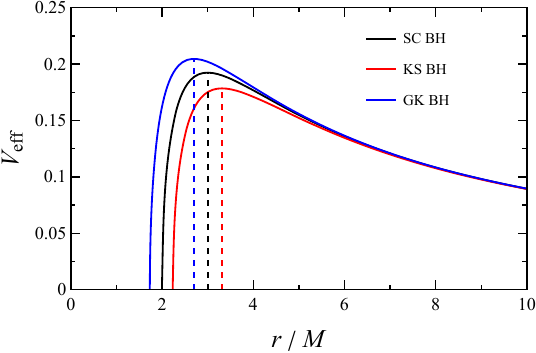}
        \caption{
        The effective potential $V_{\mathrm{eff}}$ as a function of the radial coordinate $r$ for three BHs. The black, red, and blue solid lines correspond respectively to the SC BH, the KS BH with ${a^{\mathrm{KS}}}/M=1$, and the GK BH with ${a^{\mathrm{GK}}}/M=1$.
        }
        \label{fig1}
    \end{figure}

FIG.~\ref{fig1} depicts the effective potentials for the SC BH, as well as the KS BH {with $a^{\mathrm{KS}}/M=1$} and GK BH {with $a^{\mathrm{GK}}/M=1$}. As shown in the figure, the three effective potentials start from zero at the event horizon ($f(r_{h}) = 0$), rapidly increase to their maximum values as $r$ increases, and then gradually decrease to tend toward the same value.
Compared to the SC BH, the parameter $a^{\mathrm{KS}}$ causes an overall decrease in the effective potential of the KS BH, indicating that photons require less energy to escape the BH. Instead, the parameter $a^{\mathrm{GK}}$ leads to an overall increase in the potential for the GK BH, implying that photons need more energy to escape the BH.
Corresponding to the extremum points $r_p$ of the effective potential, the KS BH has a larger $r_p$ than the SC BH, whereas the GK BH has a smaller $r_p$.
The extremum point $r_p$ of the effective potential $V_{\mathrm{eff}}$ is the radius of the photon sphere.
The photon sphere radius of the BH can be obtained by setting $\partial V_{\mathrm{eff}}/\partial r=0$.
To investigate the effects of the ``hair'' parameters of different minimally deformed SC BHs on photon spheres and shadows, we set the BH's mass as $M=1$. The photon sphere radii for the three BHs are thus

\begin{itemize}
    \item SC BH
\end{itemize}
    \begin{equation}
        r^{\mathrm{SC}}_p=3,\label{2}
    \end{equation}
\begin{itemize}
    \item KS BH
\end{itemize}
    \begin{equation}
        r^{\mathrm{KS}}_p=\sqrt{\dfrac{3}{2}(a^2+\sqrt{2{(a^{\mathrm{KS}})}^2+9}+3)},\label{3}
    \end{equation}
\begin{itemize}
    \item GK BH
\end{itemize}
    \begin{equation}
        r^{\mathrm{GK}}_p=\sqrt{\chi-{(a^{\mathrm{GK}})}^2+\dfrac{9-2{(a^{\mathrm{GK}})}^2}{\chi}+3},\label{4}
    \end{equation}
where
\begin{equation*}
\chi=\left(\dfrac{{(a^{\mathrm{GK}})}^4+{(a^{\mathrm{GK}})}^3\sqrt{{(a^{\mathrm{GK}})}^2-4}-18{(a^{\mathrm{GK}})}^2+54}{2}\right)^{1/3}.
\end{equation*}

    \begin{figure*}[hbpt]
        \centering
        \includegraphics[width=0.45\textwidth]{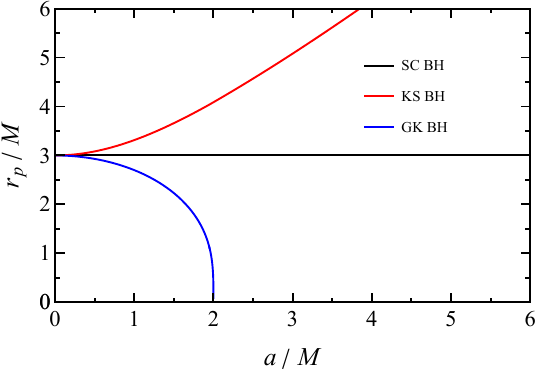}
        \hspace{10pt}
        \includegraphics[width=0.45\textwidth]{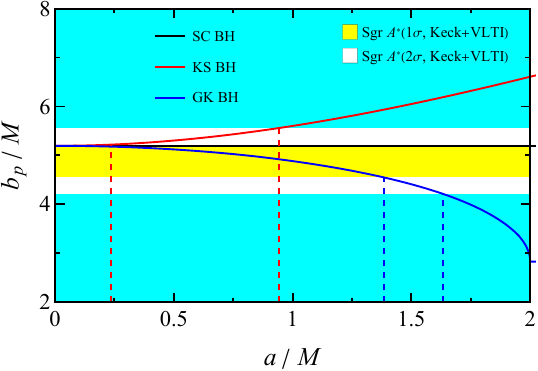}
        \caption{Left panel: Dependence of the photon sphere radius $r_p/M$ on the BH hair ${a^{\mathrm{KS}}/M}$ and ${a^{\mathrm{GK}}/M}$. As the parameter ${a^{\mathrm{KS}}/M}$ and ${a^{\mathrm{GK}}/M}$ increases, the photon sphere radius of the KS BH increases rapidly, but that of the GK BH decreases more rapidly, with the photon sphere disappearing at ${a^{\mathrm{GK}}}=2M$. Right panel: Dependence of the critical impact parameter $b_p/M$ on the BH hair ${a^{\mathrm{KS}}/M}$ and ${a^{\mathrm{GK}}/M}$. After averaging the Keck and VLTI mass-to-distance ratio priors for Sgr A*, the yellow and white regions are consistent with the EHT angular size images of Sgr A* at $1\sigma$ and $2\sigma$ levels, respectively. Conversely, the cyan region is excluded by the same observations at more than $2\sigma$.}
        \label{fig2}
    \end{figure*}

Substituting Eqs.~\eqref{2}, \eqref{3} and \eqref{4} into $b_p=1/$
$V_{\mathrm{eff}}(r_p)$ gives the critical impact parameters $b_p$ for the three BHs as follows

\begin{itemize}
    \item SC BH
\end{itemize}
    \begin{equation}
        b^{\mathrm{SC}}_p=3\sqrt{3},
    \end{equation}
\begin{itemize}
    \item KS BH
\end{itemize}
    \begin{equation}
        b^{\mathrm{KS}}_p=\dfrac{\sqrt[4]{(3{(a^{\mathrm{KS}})}^2+3\sqrt{2{(a^{\mathrm{KS}})}^2+9}+9)^3}}{\sqrt{2\sqrt{{(a^{\mathrm{KS}})}^2+3\sqrt{2{(a^{\mathrm{KS}})}^2+9}+9}-4\sqrt{2}}},
    \end{equation}
\begin{itemize}
    \item GK BH
\end{itemize}
    \begin{equation}
        b^{\mathrm{GK}}_p=\dfrac{\sqrt{\chi-{(a^{\mathrm{GK}})}^2(2/\chi+1)+9/\chi+3}}{\sqrt{1-2/\sqrt{\chi+(9-2{(a^{\mathrm{GK}})}^2)/\chi+3}}}.
    \end{equation}

FIG.~\ref{fig2} presents the dependences of the photon sphere radius $r_p$ and the critical impact parameter $b_p$ on the minimum deformation parameter ${a^{\mathrm{KS}}/M}$ and ${a^{\mathrm{GK}}/M}$ in the two deformations of SC BH.
In the left panel, as ${a^{\mathrm{KS}}/M}$ and ${a^{\mathrm{GK}}/M}$ increases from 0, the photon sphere radius of the KS BH rapidly expands from that of the SC BH, but the GK BH's photon sphere radius decreases more abruptly (vanishing at ${a^\mathrm{GK}/M=2}$).
This indicates that increasing the magnetic charge ${a^{\mathrm{GK}}/M}$ in the GK BH leads to extremal BHs or naked singularities, which do not support photon spheres and thus produce no shadows.
In the right panel, the latest EHT observations of Sgr A* \cite{r22,r39} constrain the parameters of KS and GK BHs.
The constraint data from Keck and VLTI are as follows
\begin{equation*}
        \begin{cases}
        4.55\lesssim b_p/M\lesssim5.22,&\text{within $1\sigma$},\\[1em]
        4.21\lesssim b_p/M \lesssim5.56,&\text{within $2\sigma$}.
        \end{cases}
\end{equation*}
This yields upper limits: ${a^\mathrm{KS}}/M\lesssim0.23$ ($1\sigma$) and ${a^\mathrm{KS}}/M\lesssim0.94$ ($2\sigma$) for the KS BH, and ${a^\mathrm{GK}}/M\lesssim1.38$ ($1\sigma$) and ${a^\mathrm{GK}}/M\lesssim1.63$ ($2\sigma$) for the GK BH. Consistently, EHT data strongly constrain theories where critical impact parameters increase with ${a^{\mathrm{KS}}/M}$ and ${a^{\mathrm{GK}}/M}$ \cite{r39}.
It should be emphasized that the constrained upper limits of $a^\mathrm{GK}/M$ are significantly lower than the extremal limit of the GK BH solution ($a^\mathrm{GK}/M=2$). All parameter values within the $1\sigma$ and $2\sigma$ constrained ranges have complete event horizons, which are physically self‑consistent and do not degenerate into extremal BHs or naked singularities This ensures that all the solutions studied in this paper satisfy the physical premise of BH shadow formation.

Given the monotonic relationships between event horizons, photon sphere radii, critical impact parameters, and BH "hair" parameters for KS and GK BHs, the following analysis fixes ${a^{\mathrm{KS}}/M=1}$ and ${a^{\mathrm{GK}}/M=1}$ for both hairy BHs to explore their physical differences from the SC BH.

We then focus on photon trajectories. Taking the square root of Eq.~\eqref{5} and multiplying by $\mathrm{d}\lambda/\mathrm{d}\varphi$, we obtain the ray trajectory equation
    \begin{equation}
        \dfrac{\mathrm{d}r}{\mathrm{d}\varphi}=\pm \dfrac{r^2}{b}\sqrt{1-f(r)\dfrac{b^2}{r^2}}.\label{6}
    \end{equation}
The positive/negative sign in the above equation depends on the direction of light deflection and the variation of the radial coordinate. By introducing a new variable $u\equiv1/r$, Eq.~\eqref{6} can be transformed into
    \begin{equation}
        \dfrac{\mathrm{d}u}{\mathrm{d}\varphi}=\mp \sqrt{\dfrac{1}{b^2}-u^2f(u)},\label{24}
    \end{equation}
where the negative/positive sign in this equation corresponds to the positive/negative sign in Eq.~\eqref{6}. For the three BHs under consideration, the differential equations governing the ray trajectories are

\begin{itemize}
    \item SC BH
\end{itemize}
    \begin{equation}
        \left(\dfrac{\mathrm{d}u}{\mathrm{d}\varphi}\right)^{\mathrm{SC}}=\mp \sqrt{2Mu^3-u^2+\dfrac{1}{b^2}},
    \end{equation}
\begin{itemize}
    \item KS BH
\end{itemize}
    \begin{equation}
        \left(\dfrac{\mathrm{d}u}{\mathrm{d}\varphi}\right)^{\mathrm{KS}}=\mp \sqrt{2Mu^3-u^2\sqrt{1-{(a^{\mathrm{KS}})}^2u^2}+\dfrac{1}{b^2}},
    \end{equation}
\begin{itemize}
    \item GK BH
\end{itemize}
    \begin{equation}
        \left(\dfrac{\mathrm{d}u}{\mathrm{d}\varphi}\right)^{\mathrm{GK}}=\mp \left(\dfrac{2Mu^3}{\sqrt{1+{(a^{\mathrm{GK}})}^2u^2}}-u^2+\dfrac{1}{b^2}\right)^{1/2}.
    \end{equation}

\begin{figure*}[hbpt]
        \centering
        \includegraphics[width=0.3\textwidth]{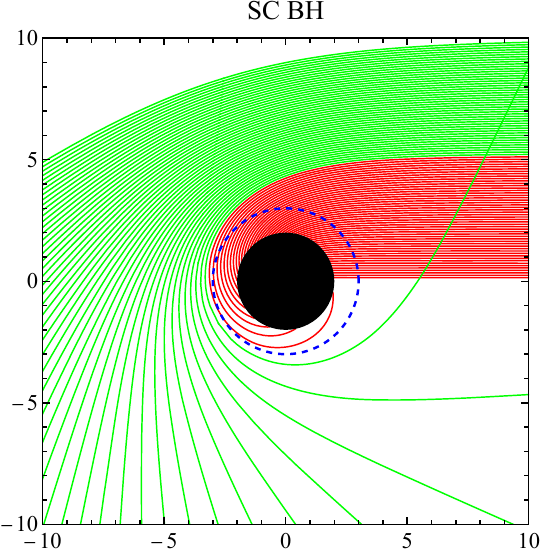}
        \hspace{10pt}
        \includegraphics[width=0.3\textwidth]{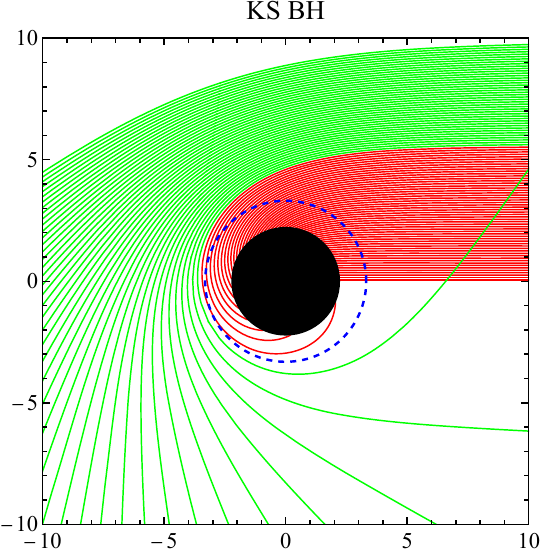}
        \hspace{10pt}
        \includegraphics[width=0.3\textwidth]{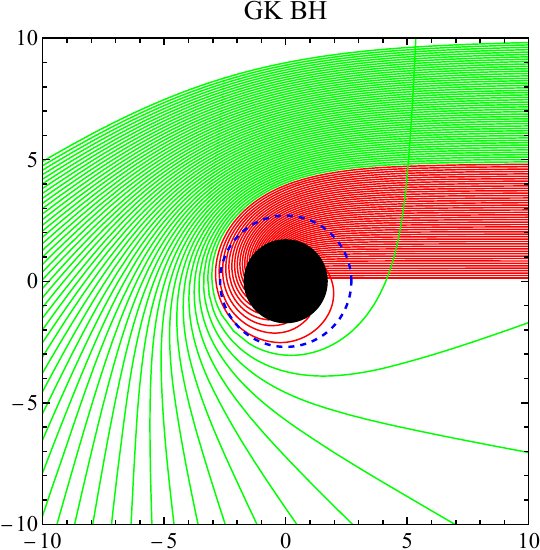}
        \caption{Photon trajectories for three BHs in the equatorial plane. From left to right: SC BH, KS BH with ${a^{\mathrm{KS}}}/M=1$, and GK BH with ${a^{\mathrm{GK}}}/M=1$. In each trajectory plot, the red and green lines represent light rays with $b<b_p$ and $b>b_p$, respectively, but the blue dashed line denotes the photon sphere. The BH is depicted as a black disk. The coordinates are scaled by $M$.}
        \label{fig3}
    \end{figure*}

Using a ray-tracing code, FIG.~\ref{fig3} plots the ray trajectories for the SC BH, the KS BH with ${a^{\mathrm{KS}}}/M=1$, and the GK BH with ${a^{\mathrm{GK}}}/M=1$.
It is evident that the minimal deformation of the SC BH endows the BH with ``hair'', thereby altering the event horizon and photon sphere.
As concluded previously, compared to the SC BH, the event horizon radius and photon sphere radius of the KS BH increase, but those of the GK BH decrease.
The weakened gravitational field of the KS BH leads to relatively insignificant light deflection in its vicinity, whereas the magnetic charge of the GK BH enhances the gravitational field, causing more pronounced light deflection.

According to the no-hair theorem, BH properties are uniquely determined by mass, charge, and angular momentum \cite{r67,r68}.
However, the regular KS and GK BHs violate the dominant energy condition, thereby extending beyond the classical no-hair theorem framework.
These solutions not only resolve the central singularity but also demonstrate that BH shadow formation critically depends on the ``hair'' parameter ${a^{\mathrm{KS}}}$ and ${a^{\mathrm{GK}}}$.
Specifically, the shadow of the KS BH expands due to enhanced photon capture, while the GK BH's shadow contracts under identical conditions.

\section{Shadows and photon rings of KS and GK BHs in different spherical accretion models\label{sub4}}
Matter in the universe, such as plasmas, gases, and dust, exists in a state of free motion.
When approaching a BH, this matter gradually accretes onto the BH under its gravitational pull, forming an accretion flow.
The accreting matter emits light during its motion, which plays a crucial role in enabling observers to detect optical phenomena such as BH shadows and photon rings.
By studying the characteristics of BH shadows and photon rings under different types of spherical accretion (static and infalling), we aim to deeply understand the impact of the accretion process on the optical manifestations of BHs and further explore the properties of BHs.
In this section, assuming luminous matter permeates the universe, we analyze how BH shadow size, brightness, and photon ring morphology depend on BH types and kinematic properties of accreting matter under these accretion conditions.

We consider an optically thin spherical accretion flow outside the BH's event horizon, either statically distributed or infalling.
The specific intensity $I_{\text{obs}}(b, \nu_o)$ observed by an observer at frequency $\nu_{o}$ can be written as \cite{r48,r70,r71,r72}
    \begin{equation}
        I_{\mathrm{obs}}(b,\nu_o)=\int_{\mathrm{ray}}{g^3j(\nu_e)}\mathrm{d}l.\label{7}
    \end{equation}
Here, $g$ denotes the ratio of the photon frequency measured by an observer at infinity $\nu_o$ to that measured by a static observer at the accretion region $\nu_e$ (i.e., $g:=\nu_o/\nu_e$), referred to as the redshift factor.
Next, $j(\nu_{e})$ is the emissivity per unit volume at frequency $\nu_e$ as measured by a static observer at the accreting matter location, and $\mathrm{d}l$ is the proper distance along the null geodesic orbit in the accretion flow's reference frame.
Finally, since the accretion flow extends to infinity, the integration path ``ray'' denotes the entire trajectory of light.
The integrated intensity is derived by integrating the specific intensity over all frequencies~\cite{r53,r73}
    \begin{equation}
        F_{\mathrm{obs}}(b)=\int_0^{+\infty}{I_{\mathrm{obs}}(b,\nu_o)}\mathrm{d}\nu_o.\label{8}
    \end{equation}

We consider a simplified monochromatic emission case in reference~\cite{r48}, where the radial distribution of the emissivity scales radially as $1/r^2$ and the rest-frame frequency is $\nu$, i.e.,
    \begin{equation}
        j(\nu_e) \propto \dfrac{\delta(\nu_e-\nu)}{r^2}.\label{9}
    \end{equation}
Denoting the 4-velocity of the accreting matter as $Z^\mu$, as well as the 4-velocity of a static observer at infinity as $U^\mu=(1,0,0,0)$. For the static accretion case, one finds
    \begin{equation}
        Z^\mu=\left(\dfrac{1}{\sqrt{f(r)}},0,0,0\right),\label{10}
    \end{equation}
while according to the constraint $g_{\mu\nu}Z^\mu Z^\nu=-1$ on the 4-velocity of the accreting matter for the free fall case, we gain
    \begin{equation}
        Z^\mu=\left(\dfrac{\mathscr{E}}{f(r)},-\sqrt{\mathscr{E}^2-f(r)},0,0\right),\label{11}
    \end{equation}
where the negative sign for the radial coordinate in the above equation is adopted considering that the accreting matter approaches the BH.
Here, $\mathscr{E}$ is the conserved energy along the geodesic of the accreting matter, given by $\mathscr{E}=-g_{\mu\nu}Z^\mu \xi_1^\nu$ and determined by the initial rest position at the onset of infall.
If the accreting matter starts falling from rest at $r = r_f$, the coordinates of $Z^\mu$ at this moment are $(1/\sqrt{f(r_f)},0,0,0)$, so $\mathscr{E}=\sqrt{f(r_f)}$.
The redshift factor $g$ can be expressed as
    \begin{equation}
        g=\dfrac{\nu_o}{\nu_e}=\dfrac{(-K^\mu U_\mu)|_{\mathrm{obs}}}{(-K^\nu Z_\nu)|_{\mathrm{em}}}.\label{12}
    \end{equation}
Here the subscript "$\mathrm{obs}$" labels the observation point in spacetime, while "$\mathrm{em}$" designates the photon emission point from accreted matter.
Finally, the proper distance $\mathrm{d}l$ is defined as \cite{r51,r70,r71}
    \begin{equation}
    \mathrm{d}l:=-K_\mu Z^\mu \mathrm{d}\lambda. \label{13}
    \end{equation}

\subsection{Shadows and photon rings of static spherical accretion}

Applying Eqs.~\eqref{17}, \eqref{7}, \eqref{8}, \eqref{9}, \eqref{10}, \eqref{12}, and \eqref{13}, the expression for the integrated intensity in static spherical accretion can be derived as
    \begin{equation}
        F_{\mathrm{obs}}(b)=\int_{\mathrm{ray}}\dfrac{f(r)^{2}}{r^2}\sqrt{\dfrac{1}{f(r)}+\dfrac{b^2}{r^2-b^2f(r)}}\mathrm{d}r.
    \end{equation}
The integrated intensities for the three BHs are
\begin{widetext}
    \begin{itemize}
    \item SC BH
\end{itemize}
    \begin{equation}
        F^{\mathrm{SC}}_{\mathrm{obs}}(b)=\int_{\mathrm{ray}}\dfrac{(r-2M)^{2}}{r^{4}}\sqrt{\dfrac{r}{r-2M}+\dfrac{b^2r}{r^3-b^2(r-2M)}}\mathrm{d}r,
    \end{equation}
\begin{itemize}
    \item KS BH
\end{itemize}
    \begin{equation}
        F^{\mathrm{KS}}_{\mathrm{obs}}(b)=\int_{\mathrm{ray}}\dfrac{\left(\sqrt{r^2-{(a^{\mathrm{KS}})}^2}-2M\right)^2}{r^{2}}\sqrt{\left(\sqrt{r^2-{(a^{\mathrm{KS}})}^2}-2M\right)\left(r^3-b^2(\sqrt{r^2-{(a^{\mathrm{KS}})}^2}-2M)\right)}\mathrm{d}r,
    \end{equation}
\begin{itemize}
    \item GK BH
\end{itemize}
    \begin{equation}
    \begin{aligned}
        F^{\mathrm{GK}}_{\mathrm{obs}}(b)=\int_{\mathrm{ray}}\left(\dfrac{\sqrt{r^2+{(a^{\mathrm{GK}})}^2}-2M}{r\sqrt{r^2+{(a^{\mathrm{GK}})}^2}}\right)^2\sqrt{\dfrac{\sqrt{r^2+{(a^{\mathrm{GK}})}^2}}{\sqrt{r^2+{(a^{\mathrm{GK}})}^2}-2M}+\dfrac{b^2\sqrt{r^2+{(a^{\mathrm{GK}})}^2}}{r^2\sqrt{r^2+{(a^{\mathrm{GK}})}^2}-b^2\sqrt{r^2+{(a^{\mathrm{GK}})}^2}+2Mb^2}}\mathrm{d}r.
    \end{aligned}
    \end{equation}
\end{widetext}

\begin{figure}[t]
    \centering
    \includegraphics[width=1\linewidth]{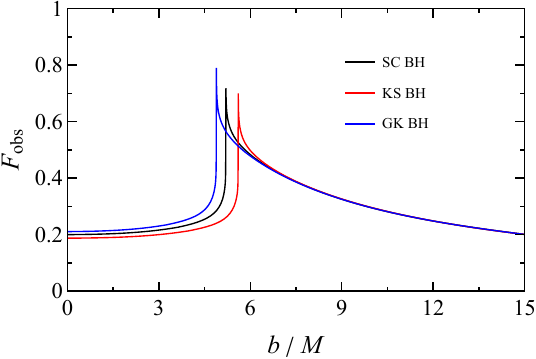}
    \caption{Distribution of the integrated intensity $F_{\mathrm{obs}}$ as a function of the impact parameter $b$ for three BHs under static spherical accretion. The black, red, and blue solid lines correspond respectively to the SC BH, the KS BH with ${a^\mathrm{KS}}/M=1$, and the GK BH with ${a^\mathrm{GK}}/M=1$. The maximum value of the integrated intensity always occurs at the critical impact parameter $b_p$ for each BH.}
    \label{fig4}
\end{figure}

FIG.~\ref{fig4} plots the distribution of the integrated intensity $F_{\text{obs}}$ versus the impact parameter $b$ for three BHs, where the parameters of both KS and GK BHs are taken as {$a^\mathrm{KS}/M=1$ and $a^\mathrm{GK}/M=1$}.
It can be seen that the integrated intensities of the three BHs all rise rapidly from a finite value, indicating that the shadow possess non-zero intensity.
When reaching the extreme point, the intensities suddenly drops and then slowly decreases to tend to the same value.
Compared with the SC BH, the impact parameter corresponding to the highest point of the integrated intensity of the KS BH is larger than the critical impact parameter of the SC BH.
The integrated intensity on the left is generally smaller than that of the SC BH, while on the right it is first larger than that of the SC BH and finally tends to the same value.
In contrast, the impact parameter corresponding to the highest point of the integrated intensity of the GK BH is smaller than that of the SC BH.
The integrated intensity on the left is larger than that of the SC BH, but on the right it is first smaller than that of the SC BH and finally tends to the same value.
Crucially, the impact parameter maximizing the integrated intensity always coincides with the critical impact parameter $b_p$, implying the photon ring radius equals $b_p$.

It is not difficult to understand the cause of such a phenomenon.
As previously known, the shadow of the GK BH contracts due to the introduction of magnetic charge, which is actually a consequence of the enhanced gravitational field.
At low $b/M$ values, gravitational lensing near the BH causes overlapping light paths, boosting the integrated intensity.
At high $b/M$, however, the intensified gravitational field reduces the photon sphere radius $r_p$.
This contracts the angular size of the shadow.
Consequently, a fixed impact parameter $b/M$ corresponds to a larger angular separation from the shadow edge, placing the region farther from the high-intensity lensing zone and thus reducing the integrated intensity.
For the same reason, quantum-corrections make the KS BH exhibit behavior opposite to that of the GK BH.

\begin{figure*}[hbpt]
    \centering
    \includegraphics[width=0.3\textwidth]{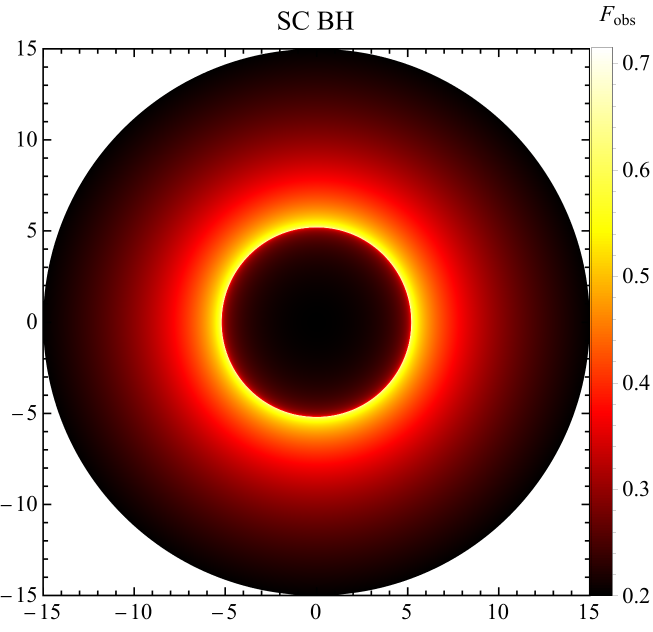}
    \hspace{10pt}
    \includegraphics[width=0.3\textwidth]{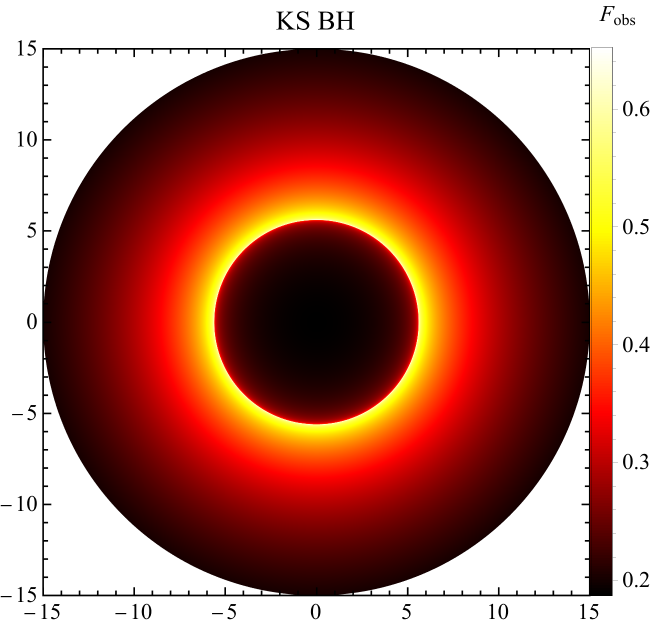}
    \hspace{10pt}
    \includegraphics[width=0.3\textwidth]{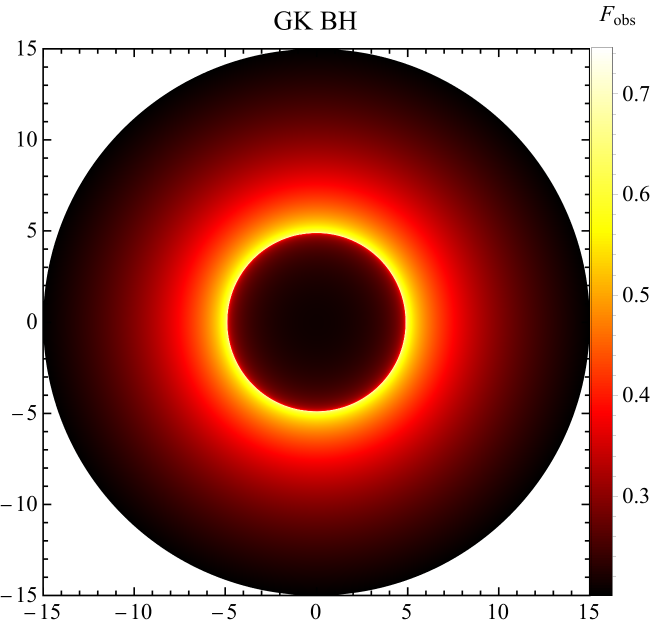}
    \caption{Optical images of BHs under static spherical accretion. From left to right are the SC BH, KS BH with ${a^\mathrm{KS}}/M=1$, and GK BH with ${a^\mathrm{GK}}/M=1$. The integrated intensity corresponding to each color is labeled on the right side of each panel, and the coordinate axes correspond to the impact parameter $b/M$.}
    \label{fig5}
\end{figure*}

Based on the integrated intensity distribution in FIG.~\ref{fig4}, FIG.~\ref{fig5} plots the optical images of the SC BH, the KS BH with ${a^\mathrm{KS}}/M=1$, and the GK BH with ${a^\mathrm{GK}}/M=1$.
It is observed that compared with the SC BH, the KS BH has a larger photon ring radius and lower shadow intensity, while the GK BH has a smaller photon ring radius and higher shadow intensity.
We have carefully analyzed the causes of the imaging results in the preceding sections.

\subsection{Shadows and photon rings of infalling spherical accretion}

By incorporating Eqs.~\eqref{11} and \eqref{12} with the condition $\mathscr{E}=1$ for matter radially infalling from infinity, we derive the redshift factor expression for spherical accretion with radial infall motion
\begin{equation}
     g_{\pm}=\dfrac{rf(r)}{r\pm\sqrt{(1-f(r))(r^2-b^2f(r))}}.
\end{equation}
In the above equation, the ``$+$'' sign represents the case where light rays are moving away from the BH, but the ``$-$'' sign represents the case where light rays are moving towards the BH. For the three types of BHs under discussion, the expressions of $g_\pm$ are as follows
\begin{widetext}
\begin{itemize}
    \item SC BH
\end{itemize}
\begin{equation}
     g^{\mathrm{SC}}_{\pm}=\dfrac{r^2-2Mr}{r^2\pm\sqrt{2Mr^3-2Mb^2r+4M^2b^2}},\label{14}
\end{equation}
\begin{itemize}
    \item KS BH
\end{itemize}
\begin{equation}
\begin{aligned}
    g_{\pm}^\mathrm{KS}=\dfrac{r\sqrt{r^2-(a^\mathrm{KS})^2}-2M}{r\pm\sqrt{\left(r-\sqrt{r^2-(a^\mathrm{KS})^2}+2M\right)\left(r^3-b^2\sqrt{r^2-(a^\mathrm{KS})^2}+2Mb^2\right)}},\label{15}
\end{aligned}
\end{equation}
\begin{itemize}
    \item GK BH
\end{itemize}
\begin{equation}
    g_\pm^\mathrm{GK}=\dfrac{r\sqrt{r^2+(a^\mathrm{GK})^2}}{r\sqrt{r^2+(a^\mathrm{GK})^2}\pm\sqrt{2M(r^2-b^2)\sqrt{r^2+(a^\mathrm{GK})^2}+4M^2 b^2}},\label{16}
\end{equation}
\end{widetext}

\begin{figure}[tb]
    \centering
    \includegraphics[width=1\linewidth]{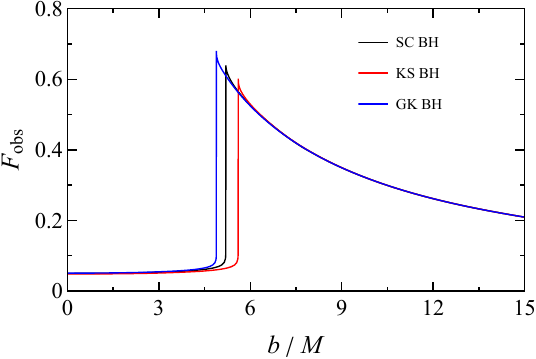}
    \caption{Distributions of the integrated intensity $F_{\mathrm{obs}}$ as a function of the impact parameter $b$ for three BHs under infalling spherical accretion. The black, red, and blue solid lines correspond respectively to the SC BH, the KS BH with ${a^\mathrm{KS}}/M=1$, and the GK BH with ${a^\mathrm{GK}}/M=1$. The maximum value of the integrated intensity always occurs at the critical impact parameter $b_p$ for each BH.}
    \label{fig6}
\end{figure}

By using Eqs.~\eqref{17}, \eqref{7}, \eqref{8}, \eqref{9} and \eqref{13}, the integrated intensity for infalling spherical accretion can be derived as
\begin{equation}
    F_\mathrm{obs}=\int_\mathrm{ray}\dfrac{g_{\pm}^3\mathrm{d}r}{r\sqrt{r^2-f(r)b^2}}.\label{18}
\end{equation}
Substituting Eqs.~\eqref{14}, \eqref{15} and \eqref{16} into Eq.~\eqref{18} gives the integrated intensity for infalling spherical accretion.
FIG.~\ref{fig6} plots the dependence of the integrated intensity $F_{\mathrm{obs}}$ on the impact parameter $b$ for three BHs under infalling spherical accretion.
It can be seen that compared with FIG.~\ref{fig4}, the variation trends of the integrated intensity in the infalling accretion and static accretion cases are similar, but the shadow intensity is significantly lower.
The extremum of each line also occurs at the critical impact parameter $b=b_p$.
The integrated intensities of the three BHs converge to similar values on the left side of $b_p$.
Compared to the static model, the integrated intensity in the infalling accretion model rises more rapidly to its maximum near $b_p$.
This indicates a greater intensity contrast across the shadow boundary in the infalling accretion model than in the static accretion model.
Similarly, compared with the SC BH, the KS BH has a larger photon ring, but the GK BH has a smaller one.

\begin{figure*}[hbpt]
    \centering
    \includegraphics[width=0.3\textwidth]{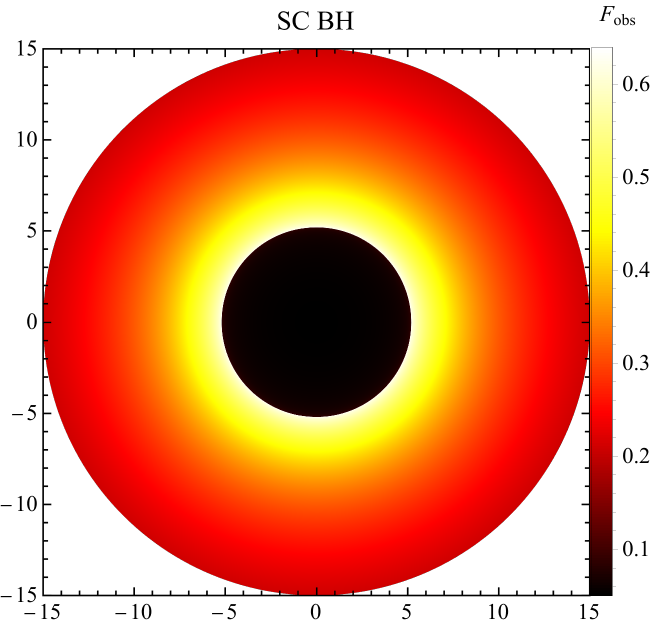}
    \hspace{10pt}
    \includegraphics[width=0.3\textwidth]{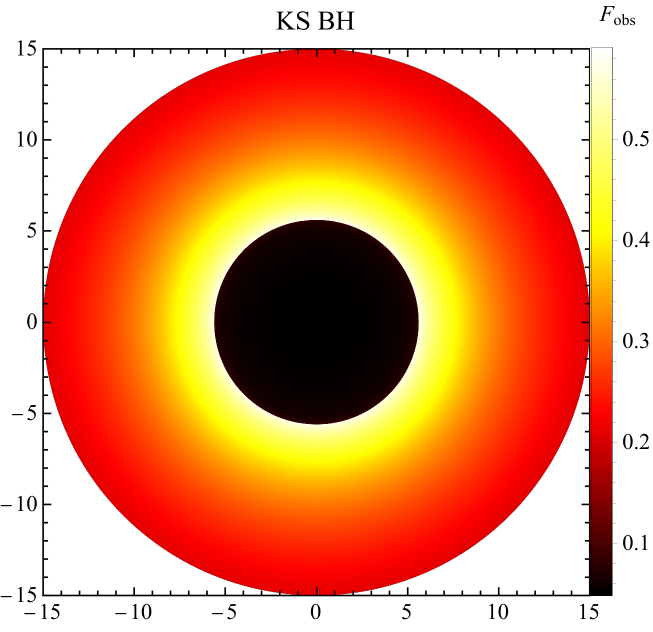}
    \hspace{10pt}
    \includegraphics[width=0.3\textwidth]{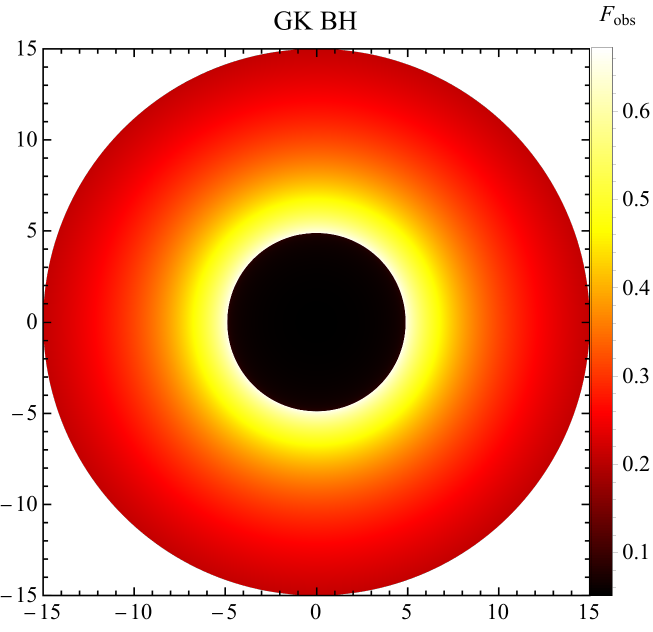}

    \caption{Optical images of three BHs under infalling spherical accretion. From left to right in the figure are the SC BH, the KS BH with ${a^\mathrm{KS}}/M=1$, and the GK BH with ${a^\mathrm{GK}}/M=1$. The integrated intensity corresponding to each color is labeled on the right side of each panel, and the coordinate axes correspond to the impact parameter $b/M$.}
    \label{fig7}
\end{figure*}

Analogously, based on FIG.~\ref{fig6}, the optical images of the SC BH, the KS BH with ${a^\mathrm{KS}}/M=1$, and the GK BH with ${a^\mathrm{GK}}/M=1$ are plotted in FIG.~\ref{fig7}. Obviously, the central low-intensity region is the shadow area, surrounded by the brightest photon ring.
As previously discussed, for BHs with the same parameters, the integrated intensity at the shadow in the static accretion case is significantly higher than that in the infalling accretion case, making the shadow of infalling accretion darker than that of static accretion.
This phenomenon originates from Doppler redshift, where the bulk infall velocity of accreting matter is antiparallel to the photon propagation direction.
Furthermore, for both minimal deformations of the SC BH, the shadow radius under spherical accretion equals the critical impact parameter $b_p$. This demonstrates that the observed shadow size is independent of the specifics of spherical accretion flows and governed exclusively by spacetime geometry.

\section{Shadows and photon rings of KS and GK BHs in prograde thin disk accretion\label{sub5}}

In this section, we will discuss disk accretion, which is abundant in the universe. In this accretion model, the disk-like accretion flow serves as the light source of the BH.
Consider a static optically thin and geometrically thin accretion disk in the equatorial plane, with the observer located directly above the disk.
The following will respectively discuss the classification of light trajectories, the transfer function, and three specific thin disk models to reveal the optical images of the thin disk for minimal deformations of the SC BH.

\subsection{Direct emission, lensed ring emission and photon ring emission\label{sub5.1}}

Light rays emitted by the disk-like accretion flow are deflected by the BH due to strong gravity, ultimately being captured by the BH or escaping to infinity.
Before reaching the observer, the light may intersect the accretion disk several times.
The azimuthal angle of light with $k$ intersections is $\varphi=\pi/2+(k-1)\pi$.
Thus, the total number of photon orbits is
\begin{equation}
    n=\dfrac{\varphi}{2\pi}=\dfrac{k}{2}-\dfrac{1}{4}.
\end{equation}
Building upon the classification framework pioneered by Gralla et al.~\cite{r53}, we adapt their criterion to categorize photons emitted from the thin accretion disk into three observable types based on orbit number $n$, with modifications specifically applied to the direct emission classification.
Specifically
\begin{itemize}
    \item When $0.25<n<0.75$, the light ray strikes the accretion disk only once, corresponding to direct emission;
\end{itemize}
\begin{itemize}
    \item When $0.75<n<1.25$, the light ray intersects the accretion disk twice, corresponding to lensed ring emission;
\end{itemize}
\begin{itemize}
    \item When $n>1.25$, the light ray intersects the accretion disk three or more times, corresponding to photon ring emission.
\end{itemize}

\begin{figure*}[hbpt]
        \centering
        \includegraphics[width=0.32\textwidth]{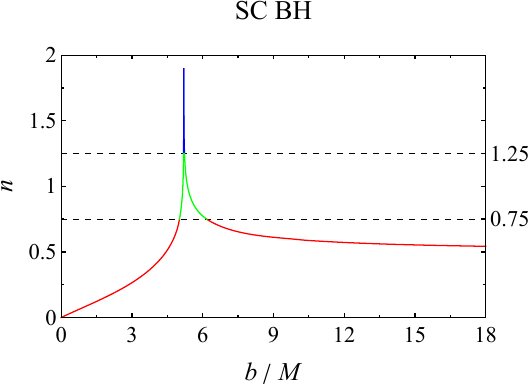}
        \includegraphics[width=0.32\textwidth]{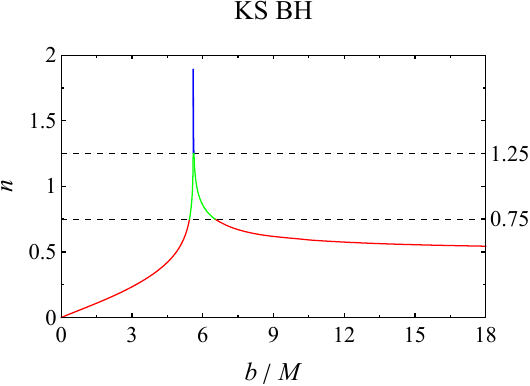}
        \includegraphics[width=0.32\textwidth]{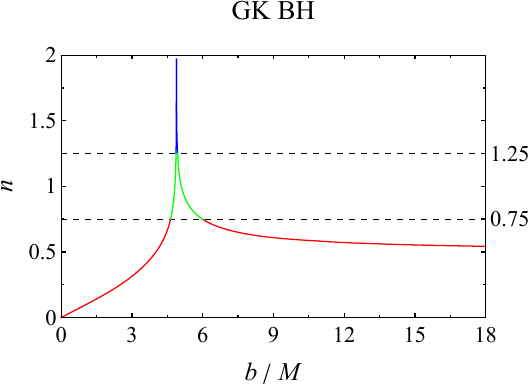}

        \caption{From left to right, the dependence of the total number of photon trajectories $n$ on the impact parameter $b$ for the SC BH, the KS BH with ${a^\mathrm{KS}}/M=1$, and the GK BH with ${a^\mathrm{GK}}/M=1$. The red, green, and blue lines correspond to direct emission, lensed ring emission, and photon ring emission, respectively.}
        \label{fig8}
\end{figure*}

FIG.~\ref{fig8} plots the relationship between the total number of photon orbits $n$ and the impact parameter $b$ for the SC BH, the KS BH with ${a^\mathrm{KS}}/M=1$, and the GK BH with ${a^\mathrm{GK}}/M=1$.
In each panel, the red, green, and blue lines correspond to direct emission, lensed ring emission, and photon ring emission, respectively.
The critical impact parameter $b_p$ corresponds to the peak of $n$, as light rays are closest to the photon sphere, where photons follow unstable circular orbits, allowing the light to undergo multiple rotations around the BH.
As b increases from zero, the total orbit number n grows for $b<b_p$, peaks at $b=b_p$, and then monotonically decreases for $b>b_p$, asymptotically approaching a constant value.
By examining the ranges of $b$ for direct emission, lensed ring emission, and photon ring emission recorded in TABLE.~\ref{tab1}, it is found that the widths of lensed ring emission and photon ring emission for the KS BH slightly shrink compared to the SC BH, while those for the GK BH slightly expand.

\begin{table*}[hbpt]
    \centering
    \normalsize
    \caption{Ranges of direct emission, lensed ring emission, and photon ring emission for the impact parameter $b$ of three BHs, where the parameters for the two minimal deformations of the SC BH are both set to {$a^\mathrm{KS}=1$ and $a^\mathrm{GK}=1$}.}

                {
    \begin{tabular}{l l l l} \hline
                     BH&  Direct emission&  Lensed ring emission& Photon ring emission\\ \hline
                     SC&  $b/M<5.015$; $b/M>6.167$&  $5.015<b/M<5.187$; $5.227<b/M<6.167$& $5.187<b/M<5.227$\\
                     KS&  $b/M<5.440$; $b/M>6.523$&  $5.440<b/M<5.596$; $5.628<b/M<6.523$& $5.596<b/M<5.628$\\
                     GK&  $b/M<4.636$; $b/M>5.986$&  $4.636<b/M<4.869$; $4.934<b/M<5.986$& $4.869<b/M<4.934$\\ \hline
    \end{tabular}
                }
    \label{tab1}
\end{table*}

\begin{figure*}[hbpt]
    \centering
    \includegraphics[width=0.3\textwidth]{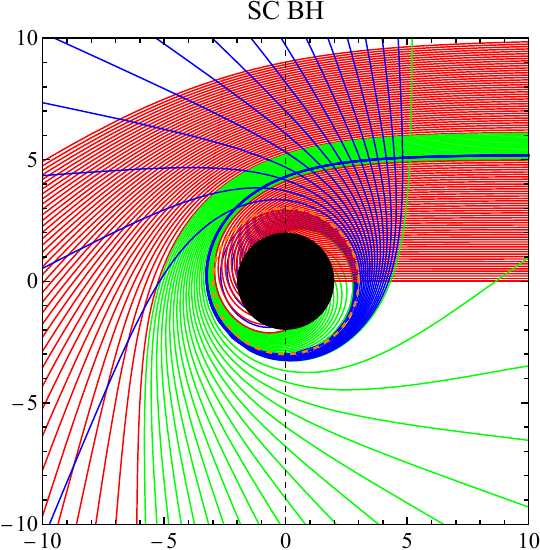}
    \hspace{10pt}
    \includegraphics[width=0.3\textwidth]{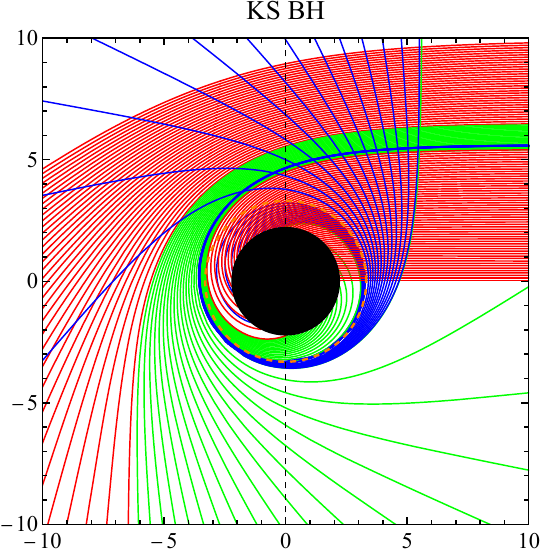}
    \hspace{10pt}
    \includegraphics[width=0.3\textwidth]{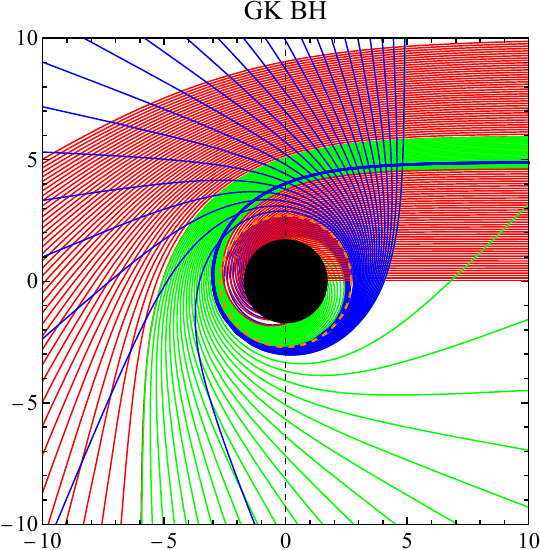}
    \caption{From left to right, the behaviors of different emissions of light around the SC BH, the KS BH with ${a^\mathrm{KS}}/M=1$, and the GK BH with ${a^\mathrm{GK}}/M=1$ are shown. Here, red lines correspond to direct emission, green lines correspond to lensed ring emission, and blue lines correspond to photon ring emission. The orange dashed line and the black disk denote the photon sphere and the BH, respectively, while the accretion disk is represented by a black dashed line. The coordinates are scaled by $M$.}
    \label{fig9}
\end{figure*}

FIG.~\ref{fig9} plots the light trajectories of different emission types for the SC BH, the KS BH with ${a^\mathrm{KS}}/M=1$, and the GK BH with ${a^\mathrm{GK}}/M=1$. Light rays of different colors represent different emission types. It should be emphasized that FIG.~\ref{fig9} is nothing more than a refinement of FIG.~\ref{fig3}. As shown, the KS BH has a smaller range of lensed rings and photon rings, but the GK BH has a larger range of such rings.

The above results are intuitive.
As concluded in Sec.~\ref{sub3}, the introduction of magnetic charge endows the GK BH with a stronger gravitational field than the SC BH.
Consequently, more light rays near the critical impact parameter $b_p$ undergo greater deflection, leading to broader lensed rings and photon rings.
Conversely, due to quantum deformation, the KS BH possesses a weaker gravitational field, resulting in fewer light rays with large deflection angles near $b_p$, and thus narrower lensed rings and photon rings.

\subsection{Integrated intensity and transfer function}

Assume that the specific intensity and frequency of light emitted by the accretion flow are $I_e(r)$ and $\nu_e$, respectively. According to Liouville's theorem, $I_e(r)/\nu_e^3$ is invariant along the entire path of light propagation, satisfying $I_e(r)/\nu_e^3=I_o(r)/\nu_o^3$, where $I_o(r)$ and $\nu_o$ are the specific intensity and frequency of the observed light.
Using Eqs.~\eqref{10} and \eqref{12}, we can gain
\begin{equation}
    I_o(r)=I_e(r)\dfrac{\nu_o^3}{\nu_e^3}=\sqrt{f^3(r)}I_e(r).
\end{equation}
To derive the integrated intensity, we need to integrate the observed specific intensity $I_o(r)$ over the entire range of the photon frequency $\nu_o$
\begin{equation}
    F_{\mathrm{obs}}(r)=\int I_o(r)\mathrm{d}\nu_o=f^2(r)I_{\mathrm{emit}},
\end{equation}
where $I_{\mathrm{emit}}$ is the emitted specific intensity, defined as
\begin{equation}
    I_{\mathrm{emit}}(r):=\int I_e(r)\mathrm{d}\nu_e.
\end{equation}

\begin{figure*}[hbpt]
    \centering
    \includegraphics[width=0.3\textwidth]{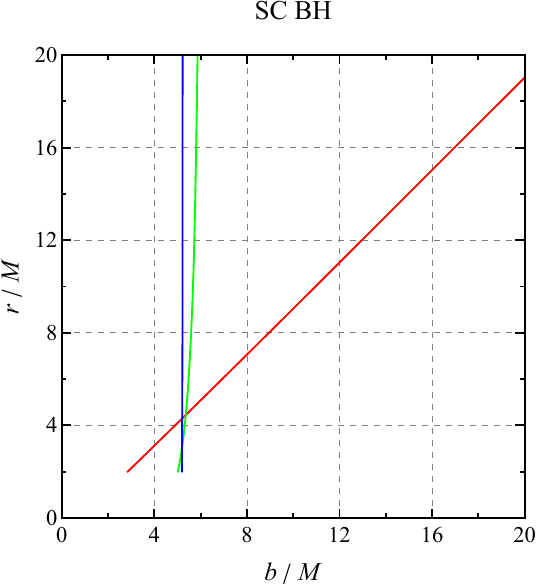}
    \hspace{10pt}
    \includegraphics[width=0.3\textwidth]{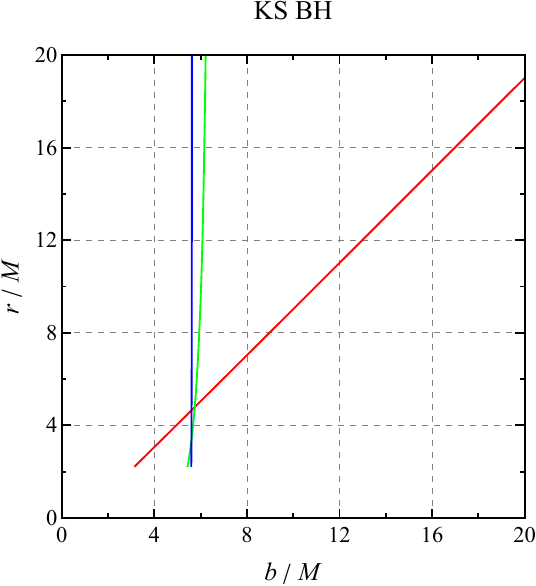}
    \hspace{10pt}
    \includegraphics[width=0.3\textwidth]{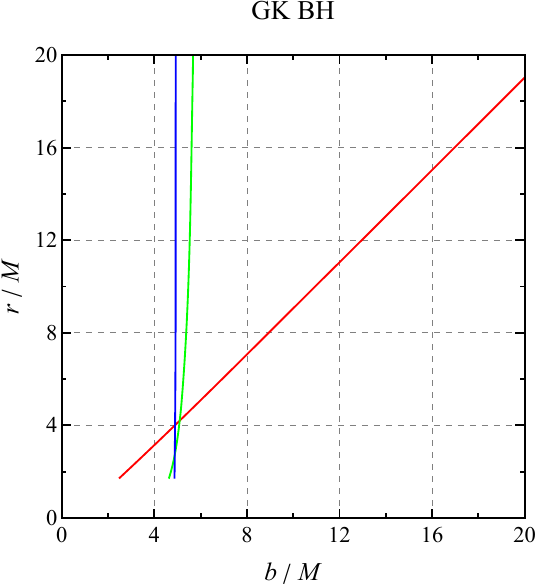}
    \caption{From left to right are the first three transfer functions of the SC BH, the KS BH with ${a^\mathrm{KS}}/M=1$, and the GK BH with ${a^\mathrm{GK}}/M=1$. The red, green, and blue solid lines represent the first, second, and third transfer functions, corresponding to direct emission, lensed ring emission, and photon ring emission, respectively.}
    \label{fig10}
\end{figure*}

In the optically thin model, each intersection of light ray with the accretion disk contributes additional specific intensity.
Therefore, the integrated intensity, which is the sum of additional intensities acquired each time the light passes through the accretion disk, should be rewritten as
\begin{equation}
    F_{\mathrm{obs}}(b)=\sum_k{f^2(r_k(b))I_{\mathrm{emit}}}(r_k(b)),
\end{equation}
where $r_k(b)$ is the transfer function \cite{r53}, representing the relationship between the photon's impact parameter and the radial coordinate of the $k$-th intersection of the light ray with the accretion disk.
This can be derived via the integral of Eq.~\eqref{24}
\begin{equation}
    r_k(b)=\dfrac{1}{u(\pi/2+(k-1)\pi,b)}.
\end{equation}
The slope of the transfer function, $\mathrm{d}r_k(b)/\mathrm{d}b=\mathnormal{\Gamma}$, is the magnification factor, governing the angular demagnification.
FIG.~\ref{fig10} plots the first three transfer functions for the SC BH, the KS BH with ${a^\mathrm{KS}}/M=1$, and the GK BH with ${a^\mathrm{GK}}/M=1$.
The red line in the figure represents the $k=1$ transfer function, corresponding to the direct emission from the accretion disk.
It is evident that the $k=1$ transfer functions of all three approximately follow a linear relationship with the impact parameter $b$ ($\mathnormal{\Gamma}\approx1$), indicating that the direct emission corresponds to the profile of a gravitationally redshifted source.
In fact, FIG.~\ref{fig10} shows that direct emission is the main component of the integrated intensity, as it accounts for the largest proportion of the radiation.

The green line denotes the $k=2$ transfer function, corresponding to lensed ring emission.
It is noteworthy that regardless of whether the BH has "hair", the slope of the second transfer function is very steep, indicating that the lensed image is highly demagnified.
Finally, the blue line near $b=b_p$ is the $k=3$ transfer function, corresponding to photon ring emission.
The slope of the third transfer function approaches infinity, implying that the photon ring image is infinitely demagnified and its flux becomes effectively negligible.

It can be seen that the shadow boundaries of all three BHs are determined by direct emission rather than photon ring or lensed ring emission.
In practice, only the first three transfer functions need to be considered at present, as higher-order emissions ($k>3$) make negligible contributions.
These results are consistent with the discussion at the end of subsection.~\ref{sub5.1}.
Notably, our findings are also consistent with the results reported by Gralla et al.~\cite{r53}.

\subsection{Observational signatures of KS and GK BHs in different thin disk models}

In the thin disk model, accretion disk serves as the primary radiation source illuminating the BH.
The emitted specific intensity of the accretion disk $I_{\mathrm{emit}}(r)$ depends not only on the radiation distribution but also on the location of the disk's inner edge $r_{\text{in}}$.
In this subsection, we intend to analyze the optical features of KS and GK BHs for three models of accretion disk emission profiles.
Notably, for simplicity, we still consider the case of monochromatic radiation.

\begin{itemize}
    \item Model I: starting from the innermost stable circular orbit
\end{itemize}

In the first model, the emission source of the accretion disk starts from the radius of the innermost stable circular orbit (ISCO) of massive particles ($r_\text{in}=r_\text{ISCO}$), and the emitted specific intensity is
\begin{equation}
    I_{\mathrm{emit}}(r)=\begin{cases}
            \exp(-r+r_\mathrm{ISCO}), & r>r_{\mathrm{ISCO}},\\[1em]
            0, & r\leq r_{\mathrm{ISCO}},
    \end{cases}
\end{equation}
where the radius of the ISCO $r_{\mathrm{ISCO}}$ is given by
\begin{equation}
    r_{\mathrm{ISCO}}=\dfrac{3f(r_{\mathrm{ISCO}})f'(r_\mathrm{ISCO})}{2f'^2(r_\mathrm{ISCO})-f(r_\mathrm{ISCO})f''(r_\mathrm{ISCO})}.
\end{equation}
Here, $f'(r)$ and $f''(r)$ denote the first and second derivatives of $f(r)$ with respect to $r$.

\begin{figure*}[hbpt]
    \centering
    \includegraphics[width=0.3\textwidth]{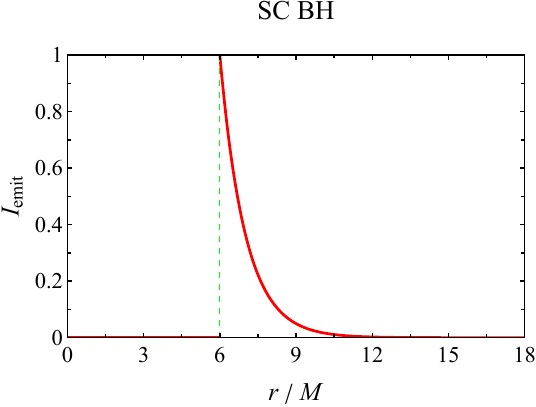}
    \hspace{10pt}
    \includegraphics[width=0.3\textwidth]{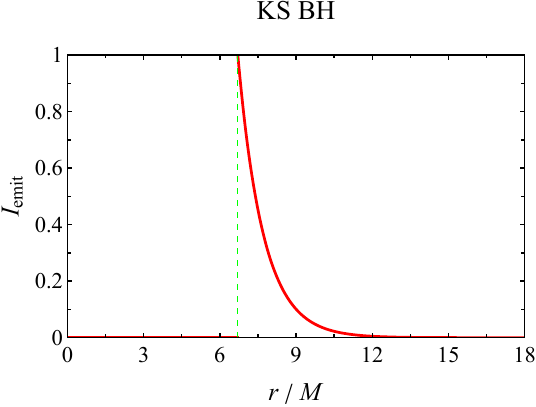}
    \hspace{10pt}
    \includegraphics[width=0.3\textwidth]{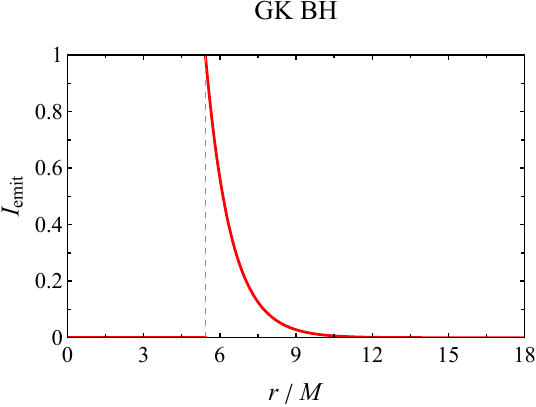}
    \includegraphics[width=0.3\textwidth]{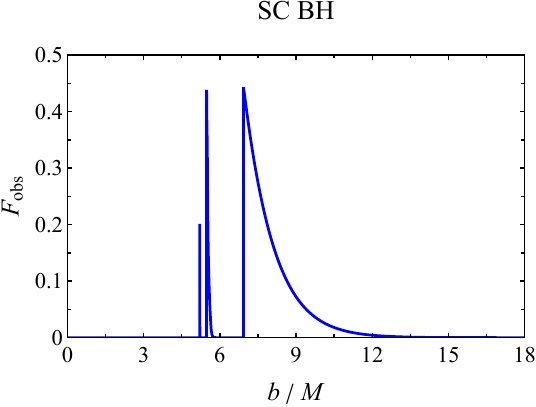}
    \hspace{10pt}
    \includegraphics[width=0.3\textwidth]{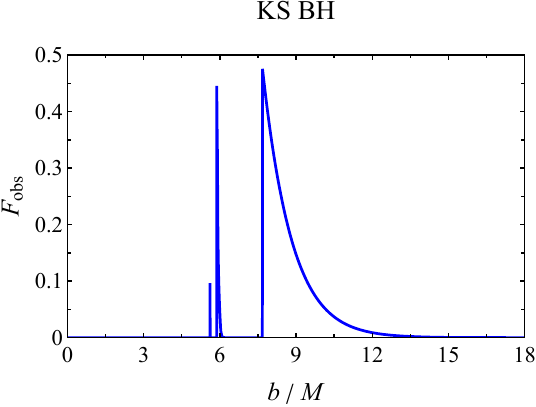}
    \hspace{10pt}
    \includegraphics[width=0.3\textwidth]{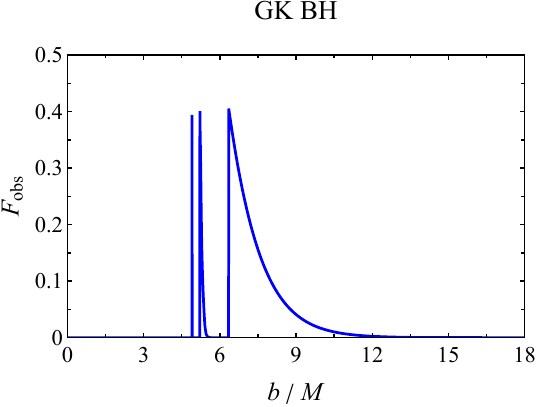}
    \includegraphics[width=0.31\textwidth]{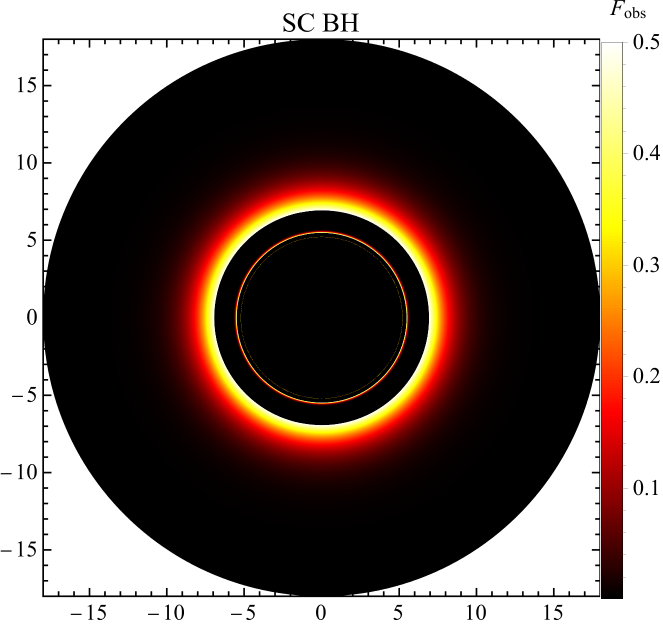}
    \hspace{10pt}
    \includegraphics[width=0.31\textwidth]{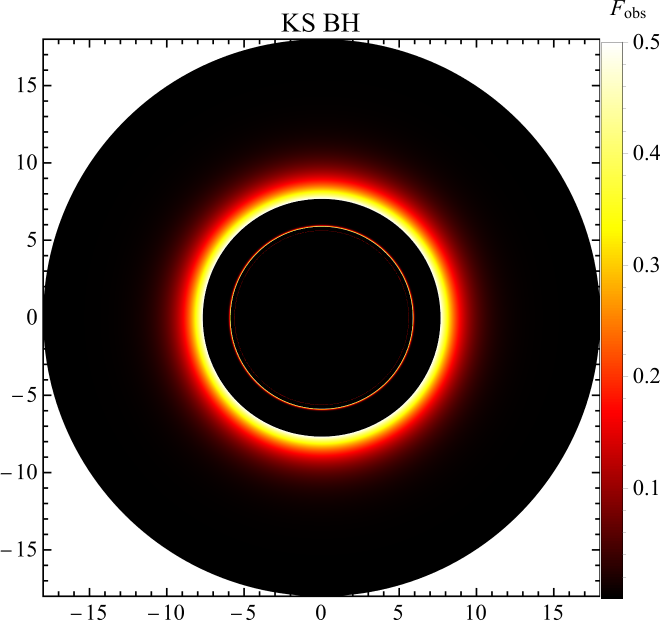}
    \hspace{10pt}
    \includegraphics[width=0.31\textwidth]{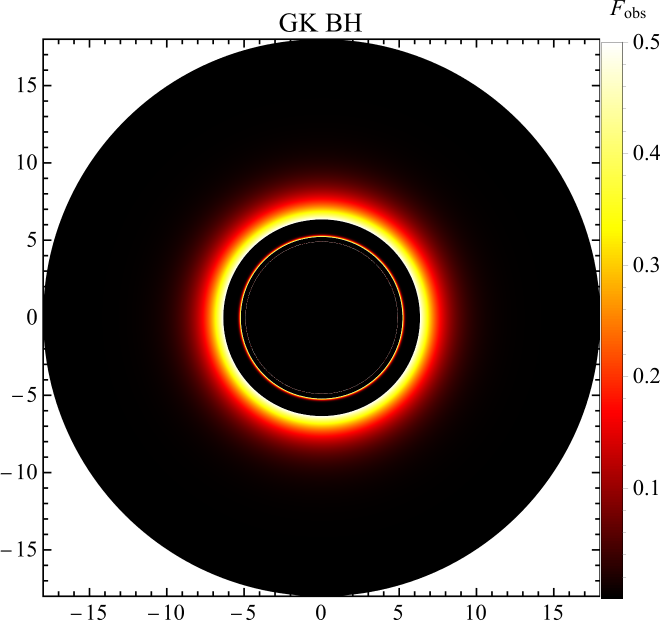}
    \caption{In the first model, $r_{\text{in}} = r_{\text{ISCO}}$. The first row shows the dependence of the emitted specific intensity $I_\mathrm{emit}$ of the thin accretion disk on the radial coordinate $r$, the second row shows the dependence of the integrated intensity $I_\mathrm{obs}$ on the impact parameter $b$, and the third row shows the optical appearance of the BH. Each column from left to right corresponds to the SC BH, the KS BH with ${a^\mathrm{KS}}/M=1$, and the GK BH with ${a^\mathrm{GK}}/M=1$. The coordinate axis corresponds to the impact parameter $b/M$.}
    \label{fig11}
\end{figure*}

FIG.~\ref{fig11} shows the emitted specific intensity of the thin disk (first row), integrated intensity (second row), and corresponding BH images (third row) for the SC BH, the KS BH with ${a^\mathrm{KS}}/M=1$, and the GK BH with ${a^\mathrm{GK}}/M=1$ in the first model.
It is evident that compared with the SC BH, the ISCO radius $r_\mathrm{ISCO}$ of the KS BH increases due to the presence of parameter {$a^\mathrm{KS}$}, but that of the GK BH decreases conversely due to {$a^\mathrm{GK}$}.
\begin{table*}[hbpt]
    \centering
    \caption{The specific positions of the lensed ring, photon ring, and initial direct emission for three BHs in the first model, where the minimum deformation parameters of the two minimal deformations of SC BHs are both taken as {$a^\mathrm{KS}/M=1$ and $a^\mathrm{GK}/M=1$}.}
                {
    \begin{tabular}{l l l l
    } \hline
                     BH&  Initial position of received direct emission  &  Lensed ring &
 Photon ring \\ \hline
        SC&  $b/M\approx6.94$&  $b/M\approx5.48$& $b/M\approx5.21$\\
        KS&  $b/M\approx7.68$&  $b/M\approx5.88$& $b/M\approx5.62$\\
        GK&  $b/M\approx6.35$&  $b/M\approx5.22$& $b/M\approx4.91$\\ \hline
    \end{tabular}
                }
    \label{tab2}
\end{table*}

From the integrated intensities of the three BHs, all exhibit three peaks.
The first peak is contributed by photon ring emission.
Notably, the photon ring intensity of the KS BH is lower ($I_\mathrm{obs}\approx0.1$), approximately half that of the SC BH, while the GK BH has a relatively higher photon ring intensity, about twice that of the SC BH.
The second peak in integrated intensity for the three BHs is contributed by lensed ring emission, with little difference among them, though the GK BH is clearly smaller than the other two.
The third peak in observed intensity and the region behind it are contributed by direct emission, which decays as $b$ increases, following a trend similar to the emitted specific intensity $I_\mathrm{emit}$.
It is evident that the peak of the KS BH is higher than that of the SC BH, but the GK BH's peak is lower.
These features are reflected in the optical images of the BHs in the third row.
From the outside to the inside of the black disk, a thin ring represents the lensed ring.
The photon ring appears inside the lensed ring but is nearly invisible due to its extremely narrow region leading to extremely low flux (observable only through magnification).
In fact, with current observational technologies, both the photon ring and lensed ring are difficult to observe, meaning that the observed shadow radius is mainly determined by direct emission.

In fact, due to the influence of different BH "hairs", the emission results of the three models are all shifted compared to the SC BH: the KS BH shifts to the right, making the starting positions of the photon ring, lensed ring, and direct emission larger; the GK BH shifts to the left, making the starting positions of the photon ring and lensed ring smaller. TABLE.~\ref{tab2} lists the specific initial positions of the photon ring, lensed ring, and direct emission for the three BHs in Model 1.
From the data in the table, the distances between the lensed ring and photon ring are similar among the three, but the distances from the photon ring to the initial position of direct emission vary significantly. Thus, using the distance between the lensed ring and photon ring $B=b_{\mathrm{L}}-b_{\mathrm{P}}$ as the basic unit, the distance from the lensed ring to the initial position of direct emission $\mathrm{\Delta}b=b_\mathrm{D}-b_{\mathrm{L}}$ can be used to distinguish different types of BHs. Specifically, this method yields $\mathrm{\Delta} b/B\approx5.41$ for the SC BH, $\mathrm{\Delta} b/B\approx6.92$ for the KS BH, and $\mathrm{\Delta} b/B\approx3.65$ for the GK BH. The results show that $\mathrm{\Delta} b/B$ is larger for the KS BH and smaller for the GK BH compared to the SC BH.

In addition, it should be emphasized that this method of discrimination is strictly established at the theoretical level, contingent upon a specific accretion disk model.
Limited by the angular resolution and observational uncertainty of the current EHT, the fine relative structure between the photon ring, lensed ring and direct emission required by this method cannot be resolved in actual astronomical observations at this stage.
In fact, this does not contradict the Kerr hypothesis supported by current EHT observations. When matching the observational precision of the EHT, the detailed features of the lensed ring and photon ring are effectively lost, and the BH shadow is instead determined by the direct emission.

\begin{figure*}[hbpt]
    \centering
    \includegraphics[width=0.3\textwidth]{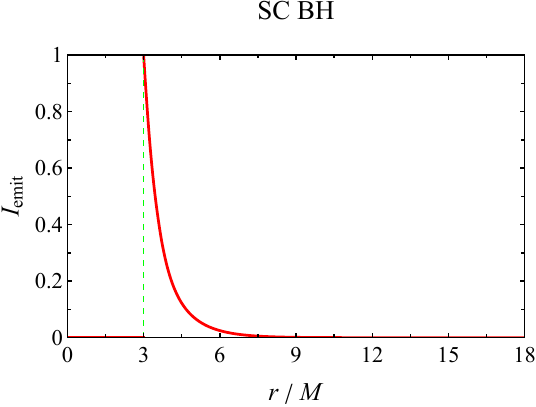}
    \hspace{10pt}
    \includegraphics[width=0.3\textwidth]{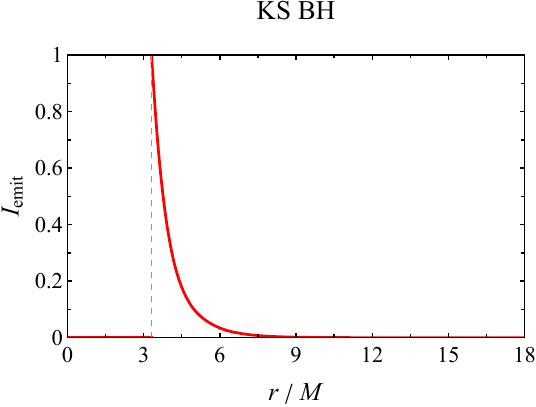}
    \hspace{10pt}
    \includegraphics[width=0.3\textwidth]{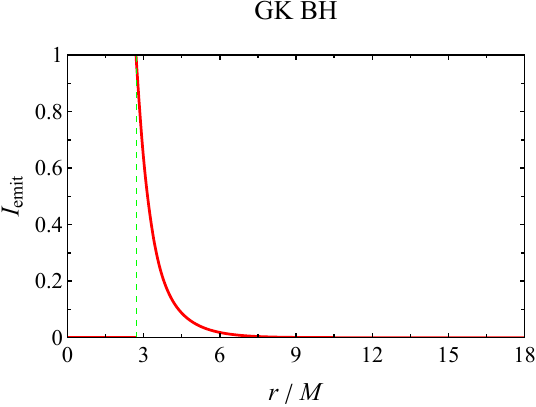}
    \includegraphics[width=0.3\textwidth]{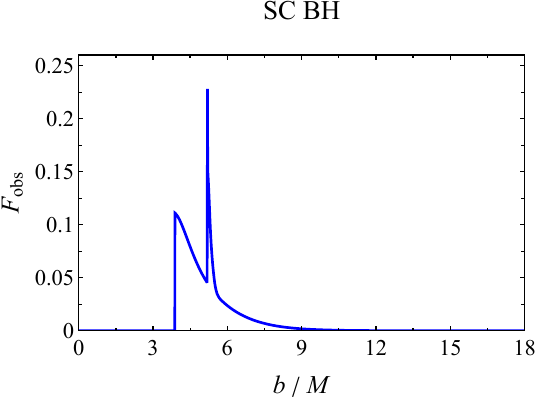}
    \hspace{10pt}
    \includegraphics[width=0.3\textwidth]{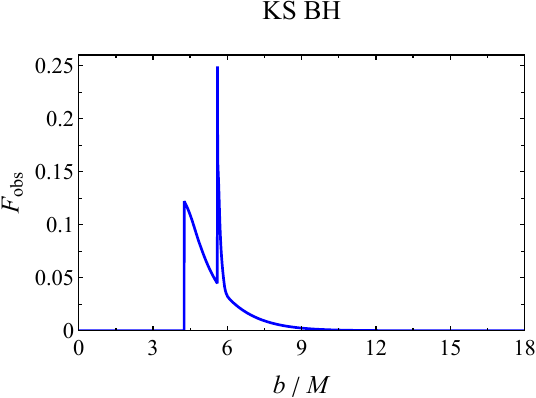}
    \hspace{10pt}
    \includegraphics[width=0.3\textwidth]{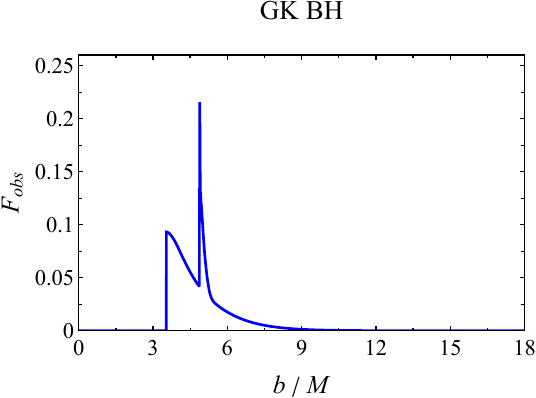}
    \includegraphics[width=0.31\textwidth]{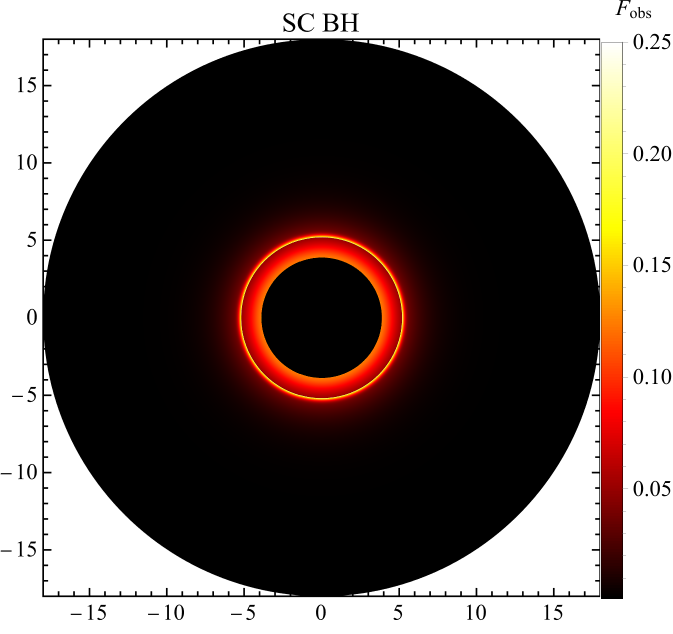}
    \hspace{10pt}
    \includegraphics[width=0.31\textwidth]{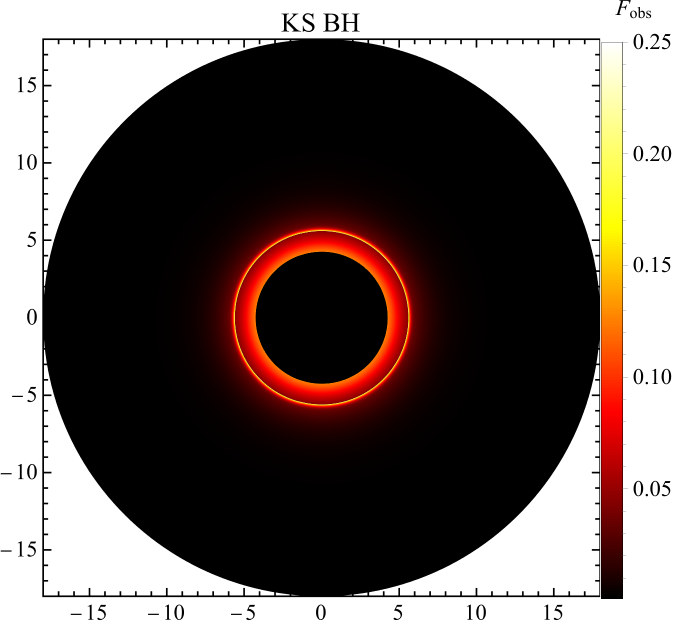}
    \hspace{10pt}
    \includegraphics[width=0.31\textwidth]{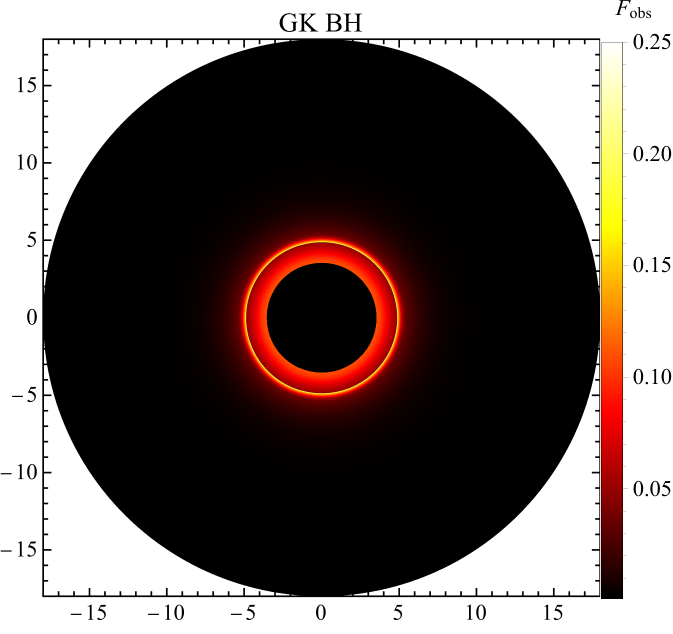}
    \caption{Similar to FIG.~$\ref{fig11}$, the model is the second model ($r_\text{in}=r_p$).}
    \label{fig12}
\end{figure*}

\begin{itemize}
    \item Model II: starting from the photon sphere
\end{itemize}

In the second model, the emission source of the accretion disk starts from the radius of the unstable circular orbit of photons ($r_\text{in}=r_p$), and the  emitted specific intensity is
\begin{equation}
    I_{\mathrm{emit}}(r)=\begin{cases}
            \dfrac{2-\tanh(r-r_p)}{2}\exp(-r+r_p), & r>r_{p},\\[1em]
            0, & r\leq r_{p}.
    \end{cases}
\end{equation}

Analogously, FIG.~\ref{fig12} shows the observational features of the three BHs in the second model.
Similar to $r_\mathrm{ISCO}$, the $r_p$ of the KS BH increases due to the presence of hair {$a^\mathrm{KS}$ and $a^\mathrm{GK}$}, while that of the GK BH decreases due to hair {$a^\mathrm{KS}$ and $a^\mathrm{GK}$} compared with the SC BH.
The integrated intensity distributions of the three BHs consistently exhibit dual-peak profiles.
The first peak corresponds to the contribution of direct emission, and the second peak corresponds to the superposition of lensed ring emission and photon ring emission, with little difference among the three.
Compared with the SC BH, the intensity of the KS BH is greater, but that of the GK BH is smaller.
In the BH images depicted in in the third row, there are two brightest rings, where the innermost ring is contributed by direct emission, and the outermost ring is contributed by the combination of lensed ring and photon ring emissions, which are difficult to distinguish in the figure.
For the shadow radius, the KS BH is larger than that of the SC BH; conversely, that of the GK BH is smaller than that of the SC BH. More importantly, the shadow radius of the second model is significantly smaller than that in the first model, which indicates that in the thin disk accretion model, the shadow radius is closely related to the details of the accretion flow.

\begin{itemize}
    \item Model III: starting from the event horizon
\end{itemize}

Finally, we consider a model where the  emitted specific intensity decays more gently than it does in the first two models.
The accretion disk radiation starts from the event horizon $r_h$, and its expression is given by
\begin{equation}
    I_{\mathrm{emit}}=\begin{cases}
                 \dfrac{\pi/2-\arctan(r-r_{\mathrm{ISCO}}+1)}{\pi/2+\arctan(r_p)},&r>r_h,\\[1em]
                 0,&r\leq r_h.
    \end{cases}
\end{equation}

\begin{figure*}[hbpt]
    \centering
    \includegraphics[width=0.3\textwidth]{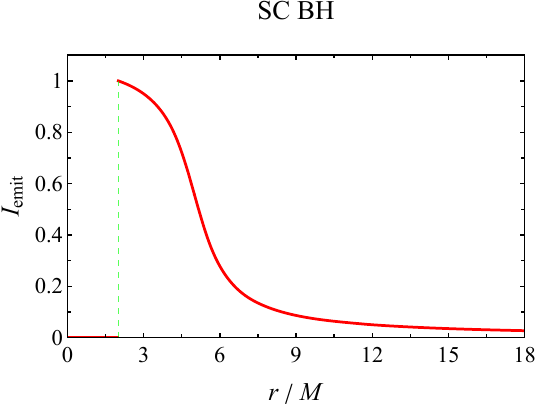}
    \hspace{10pt}
    \includegraphics[width=0.3\textwidth]{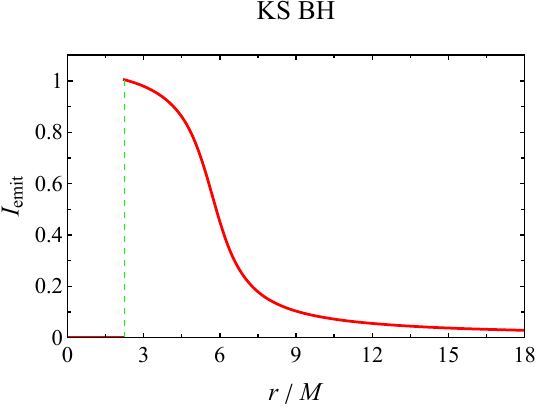}
    \hspace{10pt}
    \includegraphics[width=0.3\textwidth]{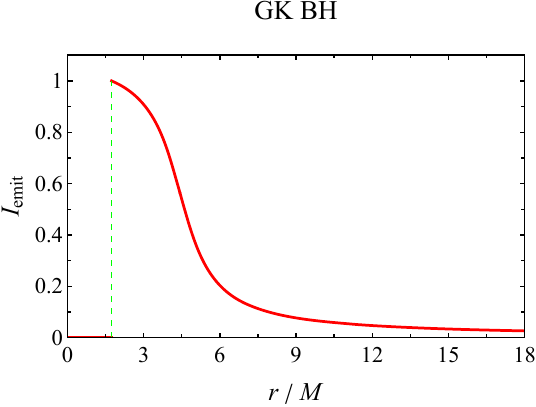}
    \includegraphics[width=0.3\textwidth]{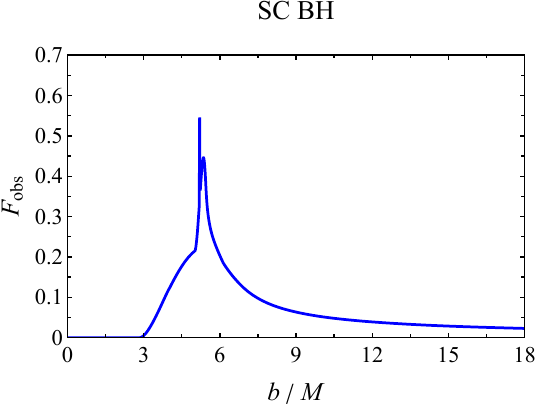}
    \hspace{10pt}
    \includegraphics[width=0.3\textwidth]{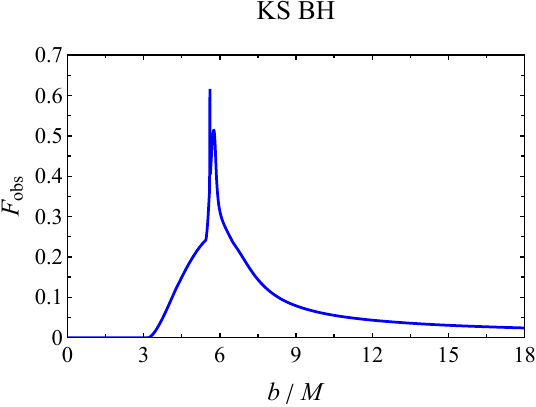}
    \hspace{10pt}
    \includegraphics[width=0.3\textwidth]{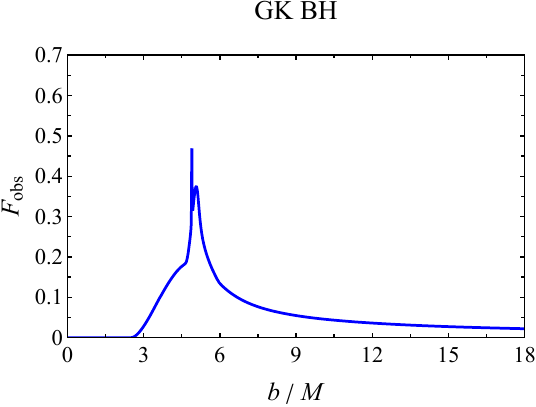}
    \includegraphics[width=0.31\textwidth]{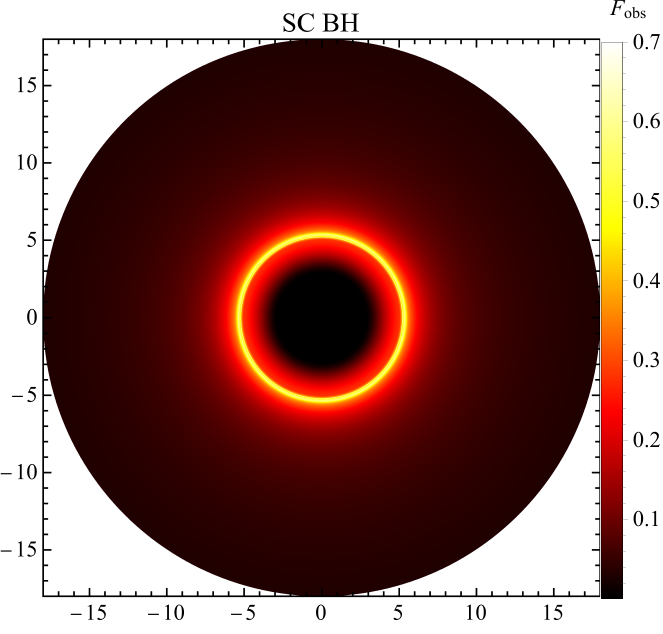}
    \hspace{10pt}
    \includegraphics[width=0.31\textwidth]{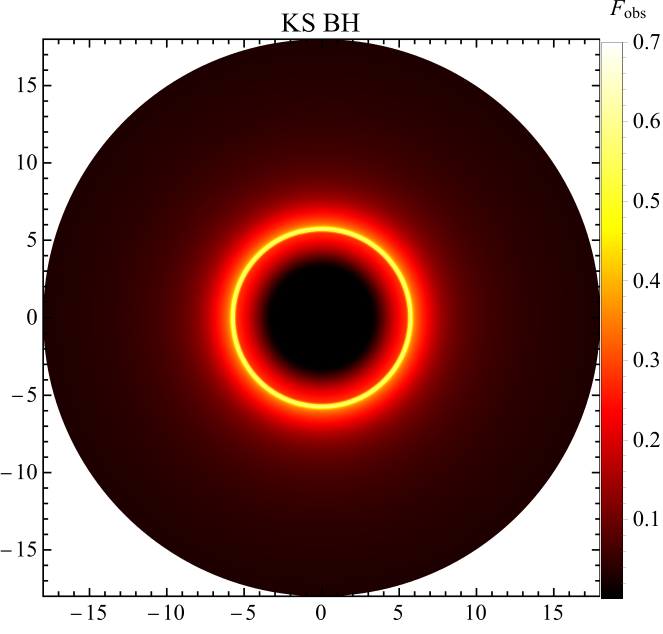}
    \hspace{10pt}
    \includegraphics[width=0.31\textwidth]{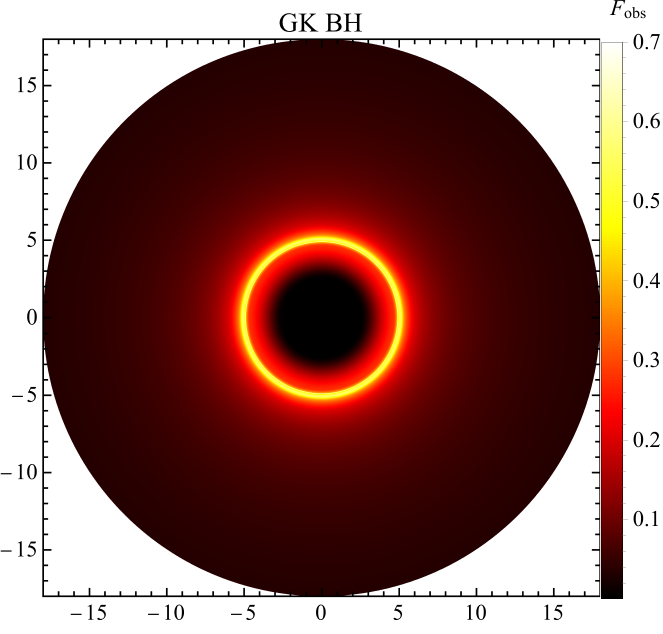}

    \caption{Similar to FIG.~$\ref{fig11}$, the model is the third model ($r_\text{in}=r_h$).}
    \label{fig13}
\end{figure*}
FIG.~\ref{fig13} shows the observational characteristics of different BHs in the third model.
Similar to the ISCO ($r_\mathrm{ISCO}$) and the photon unstable circular orbit radius ($r_p$), the event horizon radius ($r_h$) of the KS BH increases due to the presence of the hair parameter compared with the SC BH, while that of the GK BH decreases.
From the integrated intensity (second row), all three BHs exhibit two closely adjacent peaks.
The first peak is mainly contributed by photon ring emission.
Although it appears at the position of the maximum integrated intensity for all three BHs, its flux is much smaller than the nearby regions contributed by lensed ring emission and direct emission.
The photon ring of the KS BH is brighter, while that of the GK BH is dimmer.
The second peak of the integrated intensity is mainly contributed by the lensed ring.
Overall, the KS BH is brighter than the SC BH, but the GK BH is dimmer.
In the appearance of the BHs in the third row, the shadow radius $b_p$  for all three BHs slightly exceeds the critical impact parameter corresponding to the event horizon ($b=r_h$).
The bright ring outside the shadow incorporates the contributions of both the photon ring and the lensed ring, which are difficult to distinguish due to their close proximity.
The moderately bright region between the photon ring and the shadow is contributed by direct emission, indicating that the shadow boundary is formed by direct emission in this case.
It is evident that the shadow radius is smaller in this scenario.

From the above three models, it is evident that the contributions of photon ring emission and lensed ring emission are much smaller than those of direct emission, so direct emission dominates the total radiation intensity.
In Sec.~\ref{sub4}, we concluded that the observed shadow size is independent of the spherical accretion flow model, but in disk models, it is found that different disk accretion models lead to different shadow sizes.
Specifically, in the first, second, and third models discussed in this subsection, the shadow radius changes from the largest to the smallest.
The closer the accretion disk is to the BH, the smaller the starting position of direct emission becomes, and the shadow radius also becomes smaller.
Indeed, in the thin disk model, the reason why the shadow sizes of different BHs within the same model vary is the same as that in the spherical accretion model, and the discussion in the spherical accretion model can be directly applied to the thin disk model.

\section{Summary and discussions\label{sub6}}

This paper mainly studies the light trajectories, light intensity distributions, and optical appearances of two minimal deformations of the SC BH with spherical accretion and thin disk accretion, namely the KS BH and the GK BH, and discusses the deviations of them compared with the SC BH.
Due to the minimal deformations, the radii of the event horizons, photon spheres, and shadows of the hairy BHs differ from those of the SC BH.
Specifically, the KS BH has larger radius, but the GK BH has smaller radius.
These results arise because the parameters of hairy BHs alter the gravitational effects of the BHs.
Based on the shadow radius estimated by EHT, we obtained constraints on the hair parameter {$a^\mathrm{KS}$ and $a^\mathrm{GK}$} for the two BHs.
For the KS BH, the $1\sigma$ upper limit of the deformation radius $a^{\mathrm{KS}}$ is approximately 0.23, and the $2\sigma$ upper limit is approximately 0.94.
For the GK BH, the $1\sigma$ upper limit of the magnetic charge $a^{\mathrm{GK}}$ is approximately 1.38, and the $2\sigma$ upper limit is approximately 1.63.
These results can be employed to constrain the parameter selection of quantum deformations and the NED theory.

In the scenarios of two spherical accretion models, we obtain the distributions of integrated intensity for the two hairy BHs.
The results show that, compared with SC BH, KS BH leads to shadow expansion and weakens the gravitational field due to quantum-corrections, but GK BH results in shadow contraction and strengthens the gravitational field due to the introduction of magnetic charge.
Whether considering static spherical accretion or infalling spherical accretion, a stronger gravitational field leads to a brighter shadow.
It is worth noting that the details of the two spherical accretion flows do not affect the shadow radius, but they have a significant impact on the integrated intensity of the shadow. Thus, the shadow radius is only related to the spacetime structure.
Due to the Doppler effect, the shadow in the infalling spherical accretion scenario is obviously dimmer than that in the static spherical accretion scenario, and the intensity difference between the inside and outside of the shadow is larger.

In the three models of optically and geometrically thin accretion disks, we systematically investigated the observational features of these three types of BHs (SC, KS, and GK BHs).
These results show that the optical appearances of all three BHs exhibit narrow photon rings, and that lensed rings exist outside the photon rings.
Compared with the SC BH, the widths of the lensed and photon rings for the KS BH are narrower due to quantum-corrections, while those for the GK BH are broader due to the introduction of magnetic charge.
In addition, the dominant contribution to the observed integrated intensity originates from direct emission, with negligible contributions from photon and lensed rings.
Notably, distinct from spherical accretion, the boundary of the observed BH shadow in disk accretion is not determined by the critical impact parameter, photon ring, or lensed ring, but by direct emission.
Overall, among the three models, the KS BH exhibits the highest brightness, but the GK BH exhibits the lowest dimness.
In the first model, we propose a method to distinguish among the three BHs, which can aid in testing GR and other gravitational theories.

It should be emphasized that the research in this paper still has certain limitations in the context of current astronomical observations. On the one hand, the current EHT observations of Sgr A* and M87* strongly support the Kerr hypothesis, that is, the observed supermassive BHs can be fully described by the Kerr solution (SC solution for the spinless case) within the observational uncertainty.
When the deformation parameters $a^\mathrm{KS}$ and $a^\mathrm{GK}$ are within the $1\sigma$ constraint range given by EHT in this paper, the shadow radius of KS and GK BHs are highly consistent with that of the SC BH, showing significant parameter degeneracy.
On the other hand, the discrimination method for different BHs proposed in this paper relies on the measurement of the relative position of the photon ring and lensed ring, whose angular scale is far below the angular resolution limit of the current EHT. Therefore, this method is only theoretically feasible at present, and cannot be directly applied to actual astronomical observations with existing equipment.

Nevertheless, the theoretical discrimination criterion proposed in this paper still has important scientific value. With the upgrading of the next-generation EHT and the development of space very long baseline interferometry technology, the angular resolution of BH imaging will be greatly improved in the future. At that time, the method proposed in this paper can be used to test the general relativity, and search for the quantum correction effects or nonlinear electrodynamics effects in the strong gravitational field of BHs.

The conclusions of this paper are in good agreement with existing similar studies: For other quantum-corrected BHs such as 4D Gauss-Bonnet~\cite{r52} (all of which are well-defined higher-derivative gravity theories that naturally emerge in UV-complete quantum gravity like string theory), they generally possess larger horizons and photon sphere radii than SC BHs, with an expanding trend in shadow size and higher shadow intensity in accretion disk models, consistent with the behavior of KS BHs in this paper. 
In contrast, Bardeen BH, Hayward BH~\cite{b3}, and Ay\'on-Beato-Garc\'ia BH coupled with a cloud of strings~\cite{c1}, those NED BHs carrying magnetic charge, exhibit smaller horizons and photon sphere radii than SC BHs, along with shrinking shadows and lower shadow intensity in disk models, aligning with the features of GK BHs discussed herein. 
The unified imaging framework established in this work can be directly extended to comparative studies of the aforementioned analogous BHs, which is one of the core directions of our subsequent research. 
In fact, we have already found that when only scale factors (e.g., event horizon, photon sphere, and photon ring radii) are considered, Einstein cubic gravity~\cite{c2} and Einstein quartic gravity~\cite{c3} yield results identical to KS BHs in this paper, while the standalone Ay\'on-Beato-Garc\'ia BH~\cite{c4} and other NED~\cite{c5} BHs also conform to the discussion of GK BHs here. 
Therefore, one of our future works is to demonstrate that these BHs are consistent with the results of this paper in terms of shadow intensity as well.

\section*{Acknowledgements}
This work is supported by  the Guangxi Science and Technology Innovation Platform Program (Leitai Action Plan, Grant No. Guike LT2600640026), Guangxi Key R\&D Program (Guangxi Funeng Action Plan, Grant No. Guike FN2504240040). the National Natural Science Foundation of China (Grants Nos. 12105039, 12133003, and 12494574) and the Guangxi Talent Program (``Highland of Innovation Talent'').

\section*{Data availability statement}
This manuscript has no associated data. [Authors' comment: Data sharing not applicable to this article as no datasets were generated or analysed during the current study.]

\section*{Code availability statement}
Code/software will be made available on reasonable request. [Authors' comment: The code/software generated during and/or analysed during the current study is available from the corresponding author on reasonable request.]\newline

\appendix
\section{ From the 4D Einstein-Hilbert action to the 2D effective dilaton gravity theory}~\label{app}

In reference~\cite{r58}, Kazakov and Solodukhin provided a detailed derivation of the reduction from the 4D Einstein–Hilbert action to the 2D effective dilaton action. We review this procedure in appendix~\ref{app}. Kazakov and Solodukhin considered decomposing the metric $g_{\mu\nu}$ into a spherically symmetric part $g_{\mu\nu}^{\mathrm{sph}}$ and a non-spherical quantum fluctuation part $h_{\mu\nu}$, i.e.,
\begin{equation}
    g_{\mu\nu}=g_{\mu\nu}^{\mathrm{sph}}+h_{\mu\nu}.\label{a.1}
\end{equation}
To leading order, the non-spherical fluctuations $h_{\mu\nu}$ can be treated as classical quantities. 

The Einstein-Hilbert action is given by
\begin{equation}
    \mathscr{S}=\dfrac{1}{16\pi G}\int\mathrm{d}x^4\sqrt{-g}R,
\end{equation}
where $G$ is Newton's gravitational constant (we refrain from using geometric units for now), and $g$ is the determinant of the metric $g_{\mu\nu}$. Assuming that at the leading order, both the 4D metric $g_{\mu\nu}$ and the matter fields satisfy spherical symmetry. Performing a coordinate transformation $(t,r,\theta,\varphi)\mapsto(z^+,z^-,\theta,\varphi)$ on Eq.~\eqref{1} yields
\begin{equation}
    \mathrm{d}s^2=-\mathrm{e}^{\sigma(z^+,z^-)}\mathrm{d}z^+\mathrm{d}z^-+r^2(z^+,z^-)(\mathrm{d}\theta^2+\sin^2\theta \mathrm{d}\varphi^2).\label{A1}
\end{equation}
For the SC metric, $z^+$ and $z^-$ are the Kruskal coordinates. It can be shown that the Ricci scalar
\begin{equation*}
\begin{aligned}
    R=4\mathrm{e}^{-\sigma}\dfrac{\partial^2\sigma}{\partial z^+\partial z^-}-2\mathrm{e}^{-\sigma}\dfrac{\partial \ln r^2}{\partial z^+}\dfrac{\partial \ln r^2}{\partial z^-}+\dfrac{2}{r^2}+\dfrac{8}{r^2}\mathrm{e}^{-\sigma}\dfrac{\partial^2r^2}{\partial z^+\partial z^-},
\end{aligned}
\end{equation*}
has its first term identical to the Ricci scalar $R^{(2)}$ of the 2D metric
\begin{equation}
    (\mathrm{d}s^2)^{(2)}=-\mathrm{e}^{\sigma(z^+,z^-)}\mathrm{d}z^+\mathrm{d}z^-,
\end{equation}
namely,
\begin{equation}
    R=R^{(2)}-\dfrac{2}{r^2}(\nabla r)^2+\dfrac{2}{r^2}+\dfrac{2}{r^2}\square r^2,
\end{equation}
where $(\nabla r)^2=g^{(2)\alpha\beta}(\nabla_\alpha r)\nabla_\beta r$, $\square=g^{(2)\alpha\beta}\nabla_\alpha\nabla_\beta$, and $\alpha,\beta$ take values 0 or 1. Here, $g^{(2)}_{\alpha\beta}$ is the 2D metric determined by $(\mathrm{d}s^2)^{(2)}$. From Eq.~\eqref{A1}, we find $g=(\mathrm{e}^{2\sigma}r^4\sin^2\theta)/4$, and thus the action can be rewritten as
\begin{widetext}
    \begin{equation}
    \mathscr{S}=\dfrac{1}{16\pi G}\int_0^{2\pi}\mathrm{d}\varphi\int_0^{\pi}\mathrm{d}\theta\int\mathrm{d}^2z\sqrt{-g}R
    =\dfrac{1}{8G}\int \mathrm{d}^2z\left[r^2R^{(2)}-2(\nabla r)^2+2\right],\label{AS}
\end{equation}
\end{widetext}
where $\mathrm{d}^2z=\mathrm{d}z^+\mathrm{d}z^-$ and we have omitted the boundary term~$\square r^2$.

Note that Eq.~\eqref{AS} actually belongs to a specific case of 2D dilaton gravity~\cite{A1,A2,A3,A4}. Introducing the dilaton field $\phi=\ln(r^2/G)$, Eq.~\eqref{AS} can be transformed into the form of a dilaton gravity action
\begin{equation}
    \mathscr{S}=\dfrac{1}{8}\int \mathrm{d}z^2\left[\mathrm{e}^\phi\left(R^{(2)}-\dfrac{1}{2}(\nabla\phi)^2\right)+U(\phi)\right],
\end{equation}
where $U(\phi)=2/G$. Considering that quantum-corrections will alter the form of the potential function $U(\phi)$, we now introduce the dimensionless variable $\xi=r/\sqrt{G}$ and generalize the above expression to an arbitrary dilaton potential, obtaining the generalized action
\begin{equation}
    \mathscr{S}=\dfrac{1}{8}\int\mathrm{d}z^2\left[\xi^2R^{(2)}-2(\nabla \xi)^2+\dfrac{2}{G}U(\xi)\right] \, .\label{a.8}
\end{equation}
Here is Eq.~\eqref{19}.

Varying the action with respect to the 2D metric $g^{(2)\alpha\beta}$ and the field $\xi$ yields the equations of motion given in Eq.~\eqref{20}
\begin{equation}
\begin{aligned}
    2\xi\nabla_\alpha\nabla_\beta\xi=&g^{(2)}_{\alpha\beta}\left[\dfrac{1}{G}U(\xi)+2\xi\square\xi+(\nabla\xi)^2\right],\\
    0=&2\square\xi+\xi R^{(2)}+\dfrac{1}{G}\dfrac{\mathrm{d}U(\xi)}{\mathrm{d}\xi}.\label{AE}
\end{aligned}
\end{equation}
Eq.~\eqref{AE} indicates the existence of a 2D Killing vector field $\xi_\alpha=\varepsilon_{\alpha\beta}\nabla^\beta\xi$ ($\varepsilon_{\alpha\beta}$ is the 2D Levi-Civita symbol) satisfying the Killing equation
\begin{equation}
    \nabla_\alpha\xi_\beta+\nabla_\beta\xi_\alpha=0.
\end{equation}
Therefore, the field $\xi$ can be chosen as a spacelike coordinate, allowing $(\mathrm{d}s^2)^{(2)}$ to be written as
\begin{equation}
    (\mathrm{d}s^2)^{(2)}=-f(\xi)\mathrm{d}t^2+\dfrac{G}{f(\xi)}\mathrm{d}\xi^2.
\end{equation}
This form satisfies the relations
\begin{equation}
\begin{split}
    &\square \xi=\dfrac{1}{\sqrt{-g^{(2)}}}\dfrac{\partial}{\partial x^\alpha}\left(\sqrt{-g^{(2)}}g^{(2)\alpha\beta}\dfrac{\partial}{\partial x^\beta}\xi\right)=\dfrac{1}{G}\dfrac{\mathrm{d}f(\xi)}{\mathrm{d}\xi},\\[1em]
    &(\nabla\xi)^2=g^{(2)\alpha\beta}\dfrac{\partial\xi}{\partial x^\alpha}\dfrac{\partial\xi}{\partial x^\beta}=\dfrac{1}{G}f(\xi),\label{a.12}
\end{split}
\end{equation}
\newline
where $g^{(2)}$ is the determinant of the metric $g^{(2)}_{\mu\nu}$. Substituting Eqs.~\eqref{a.12} into Eq.~\eqref{AE} leads to
\begin{equation}
    f(\xi)=-\dfrac{2M}{\xi}+\dfrac{1}{\xi}\int^\xi U(\xi)\mathrm{d}\xi,
\end{equation}
where the integral is taken as indefinite and the integration constant is omitted.
Finally, reverting $\xi$ back to $r$ and employing geometric units ($G=1$) give
\begin{equation}
    f(r)=-\dfrac{2M}{r}+\dfrac{1}{r}\int^r U(r)\mathrm{d}r.\label{AKS}
\end{equation}
This is Eq.~\eqref{ks} in subsection~\ref{sub2.1}, and plugging it into Eq.~\eqref{1}, the 4D metric can be derived.

\bibliography{ref}
\end{document}